\documentclass[DIV=default, 11pt]{scrartcl} %STIX1COL,STIX2COL,STIXSMALL

\usepackage[utf8]{inputenc}
\usepackage[T1]{fontenc}
\usepackage[english=british]{csquotes}

\MakeAutoQuote{“}{”}
\MakeOuterQuote{"}
\usepackage[british]{babel}
\usepackage{graphicx}
\usepackage{tabularx}
\usepackage{booktabs}
\usepackage[stretch=10,shrink=10,
			]{microtype}
\usepackage[osf]{garamondlibre}
\usepackage[style=authoryear, date=year, isbn=false]{biblatex}
\usepackage{hyperref}
\newcolumntype{Y}{>{\arraybackslash}p{0.33\linewidth}}

\begin{document}

\title{AI-Augmented Human Resource Management?}

\subtitle{Insights from German companies}

\author{Yannick Kalff\thanks{\href{mailto:yannick.kalff@htw-berlin.de}{yannick.kalff@htw-berlin.de}. \url{https://yannickkalff.de}. ORCID: \url{https://orcid.org/0000-0003-1595-175X}.}~\,\&~Katharina Simbeck\thanks{\href{mailto:simbeck@htw-berlin.de}{simbeck@htw-berlin.de}. \url{https://iug-lab.de}. ORCID: \url{https://orcid.org/0000-0001-6792-461X}.}}

\publishers{HTW Berlin University of Applied Sciences, Berlin}

\maketitle

\renewcommand\thefootnote{}

\renewcommand\thefootnote{\arabic{footnote}}
\setcounter{footnote}{1}
\section*{Abstract}
This study examines the integration of AI into Human Resource Management in German companies. We ask if and how AI-based technologies are \enquote{augmenting} human resource management. Organisations employ generative AI or predictive analytics to transform traditional human resource functions, to streamline routine tasks and to reallocate resources toward strategic, people-centred activities. Our findings from interviews and group discussions and a survey (N=410) reveal that while AI tools enhance HR analytics capabilities, their adoption mainly serves efficiency and rationalising goals. The introduction of AI tools is shaped by organisational transformation factors such as digital infrastructure, co-determination frameworks, and ethical implications. The research highlights both the strategic potential for improved talent development and the challenges posed by data governance and algorithmic transparency. Overall, this work contributes to understanding the ambiguous role of technological change in HR, which promises to augment predictive capabilities yet serves the ends of efficiency and rationalisation.\\ Keywords: AI-augmented HRM, Generative AI, HR Analytics, Organisational Transformation, Co-determination, Rationalisation

\section{Introduction}\label{sec:intro}
AI technologies have become a pivotal driver of organisational transformation, especially in data-intensive fields like Human Resource Management (HRM) \parencite{Chowdhury.2023, Budhwar.2023, Bankins.2024}. The recent proliferation of machine learning and, notably, generative AI following the public release of ChatGPT in 2022 is reshaping workplace practices by enabling advanced data analytics, automation, and decision support systems \parencite{Budhwar.2022, Budhwar.2023}. Building on a decade of digitalisation and the \enquote{data imperative} in organisations \parencite{Fourcade.2016}, AI is labelled as the next disruptive technology to transform HRM and other organisational departments \parencite{Arora.2021, Coolen.2023}.

AI-driven tools promise to support core HRM functions including recruitment, performance evaluation, workforce planning, and talent development by providing data-driven insights, generating predictions, and automating repetitive tasks \parencite{McCartney.2022, Torre.2022}. Utilising data sources such as work processes, business intelligence, recruiting processes (i.\,e. CVs, certificates, cover letters, video interviews, interview transcripts), individual performance metrics, engagement, learning, or compensation promise deeper insights and possibilities for HRM to act. This technological shift is altering HR roles, skill requirements, and professional identities, as HR managers increasingly require analytical and data science qualities \parencite{Loscher.2022, Cayrat.2023}. 

While AI in HRM, commonly referred to as HR analytics or people analytics \parencite{Wirges.2023, Torre.2022}, offers potential benefits such as reducing human bias, efficiency gains or improved decision quality; it also brings significant risks, including algorithmic discrimination and a lack of transparency or accountability in automated decisions \parencite{Kelan.2024, Schellmann.2024, Ayling.2022, Du.2024, Narayanan.2024}. The ethical and social implications of data-driven HRM remain contested in academic and practitioner discourses \parencite{Simbeck.2019, Kochling.2021, Kochling.2024}.

However, scientific and practitioner debates regarding the actual application and utility of AI-driven tools in HRM are controversial. The urgency for HR Analytics and AI is often driven less by technological capability and more by comprehensive rationalization strategies aimed at achieving immediate efficiency gains. As such, technologies like HR Analytics serve less as disruptive innovations for work tasks and processes, and more as instruments for realising corporate efficiency objectives \parencite{Nawaz.2024}.

AI technologies manifest in two ways: on one hand, they automate repetitive mental tasks and administrative activities; on the other hand, they support and enhance human activities. As a result, AI can either replace or augment HRM functions. The concept of \textit{augmentation} is particularly noteworthy. Rather than replacing HR professionals, augmentation involves a symbiotic relationship in which AI tools support and extend the competencies of HR managers in complex areas such as recruitment, personnel development, and strategic planning \parencite{Prikshat.2023, Fenwick.2024}. Augmented HRM seeks to redefine professional boundaries by combining the strengths of human intuition, ethical judgment, and socio-emotional understanding with the efficiency and analytical capabilities of AI \parencite{Prikshat.2023a}. Ultimately, augmentation is seen as empowering HR departments to enhance their relevance and impact within organisations, leading to better outcomes for both management and employees, while also raising critical questions about professionalization, ethics, and competencies in an AI-driven context \parencite{Bohmer.2023, Diefenhardt.2024}.  \textcite{Das.2026} argue that the transformative shift in human resources (HR) incorporates a temporal dimension, evolving from the automation of administrative tasks to the augmentation of HR managers' complex, human-centred decision-making. As a result, HR is in a position to gain strategic significance within organisations and can adopt a more people-centred approach to integrating employees into the organisation \parencite{Diefenhardt.2024}. However, \textcite[194]{Raisch.2021} challenge the dichotomous \enquote{either -- or} perspective, arguing instead for an interdependent relationship between automation and augmentation.

In this context, our article examines the empirical reality of AI adoption in HRM in Germany. We contend that AI in HRM produces both automation and augmentation effects that are closely linked. In particular, we investigate the organisational and corporate context in which AI-based HRM tools are implemented and developed to achieve efficiency gains -- gains that are, in turn, perceived as enhancing the work of HR managers. Accordingly, our research focuses on three key dimensions: the tools themselves, the corporate organisational context, and the HR managers who apply these tools in their daily work. We ask the following research questions:
\begin{description}
\item[RQ1] How do HR managers perceive and respond to the ongoing transformation driven by AI?
\item[RQ2] What are the defining characteristics of organisational structures that employ AI in HRM?
\item[RQ3] In what ways do AI tools augment HRM processes and work?
\end{description}
We advance the understanding of AI’s impact on the work of HR managers and examine how HRM is evolving through the use of HR analytics. Our contribution provides a nuanced perspective on the simultaneous dynamics of automation and augmentation, demonstrating how these forces shape the specific ways HR managers use technology. Augmentation, therefore, is not the result of isolated symbiotic relationships between humans and AI, but is instead deeply embedded within the labour process and its techno-organisational structures. 

The structure of this contribution is as follows: Section~\ref{sec:stateofart} reviews the current state of research on the digitalisation trajectory of HRM, focusing on the emergence of AI and machine learning. Section~\ref{sec:methods} outlines the research design, which combines qualitative interviews, group discussions, and a quantitative survey of HR professionals to assess the scope, impact, and perception of AI-based HRM in practice. Our findings, presented in Section~\ref{sec:results}, indicate that the use of AI in German HR practices centres on automating repetitive, tedious, and low-value tasks, thereby freeing resources for interpersonal engagement and fostering a more strategic role for HR. In Section~\ref{sec:discussion}, we discuss these results and argue for a practical shift in HR work that necessitates evolving qualification profiles in response to efficiency-driven and rationalisation processes within the HR field.

\section{Technological change in HRM and HRM work}\label{sec:stateofart}
Research on AI in HRM has gained considerable momentum over the past five years, with much of the focus placed on technical perspectives -- examining various technological systems, machine learning approaches, associated tools, and their efficiency or effectiveness \parencite{Qamar.2021, Budhwar.2022, Mer.2023}. Conversely, a growing body of literature investigates the social and ethical implications of technological change in the workplace, its effects on the labour process, and broader HRM practices \parencite{Arora.2021, Qamar.2021, Weiskopf.2023}. It is important to note, however, that the adoption of AI in HRM is deeply rooted in the broader digitalisation of business processes and the increasing demand for quantifiable insights and data-driven decision-making strategies. In the following, we first review existing research on the historical development of HRM and digital technologies. We then turn to studies examining firm structure and organisational aspects of HRM applications. Finally, we discuss research addressing the impact of AI on work tasks throughout the employee lifecycle, managerial identities, and the associated ethical and legal challenges. 

\subsection{A brief history of datafication of HRM and HRM work}
%Einleitender Absatz über Aufgaben und Ziele von HRM und welche Daten hierfür anfallen/genutzt werden.
HRM has traditionally been a people-centred organisational unit, with its identity grounded in working with individuals to help them thrive in their work environment while advancing organisational goals. The professional ethos of HRM relies on intuitive and tacit knowledge, such as gut feeling and experience-based decision-making \parencite{Bolander.2013}. Recently, however, both researchers and an increasing number of practitioners have argued that decisions should be based on objective data and analytics \parencite{Korherr.2023}, and this perspective increasingly includes HRM. As a result, the processes of datafication and data analysis -- driven by ongoing digitalisation in business and its organisational units -- are viewed as positive steps towards making better decisions. Nevertheless, when it comes to analytics capabilities, HRM is often seen as \enquote{lagging behind} (\cite[9]{Angrave.2016}; \cite[14]{Coolen.2023}). 

In the 1960s and 1970s, Human Resource Management (HRM) began adopting labour-saving technologies to streamline administrative tasks and enhance operational efficiency \parencite{Kim.2021}. The introduction of personal computers further strengthened HR capabilities by facilitating the development of Human Resource Information Systems (HRIS), which enabled improved data processing within typical HR departments. HRIS supported more advanced, data-driven HR functions and standardised HR processes across both local and global organisations. 

By the 2000s, HR practices increasingly incorporated the Internet and networked technologies. This period marked the rise of e-HR and early HR analytics, which utilised statistical methods and \enquote{small} data sets -- by contrast with the current, often ambiguous notion of \enquote{Big Data} \parencite[2]{Angrave.2016}. Recruitment, training, and other HR activities began leveraging online platforms and data-driven approaches, shifting core business processes to become more data-centric and insight-driven \parencite{Coron.2022}. During the 2000s, the term "HR analytics" emerged, encompassing a range of definitions from enhanced analytical insights and decision support to scenario prediction built upon systematic use of \textit{HR metrics} \parencite{Marler.2017}. As big data and advanced analytics became integrated into HR practices -- empowered by greater computational capabilities, both on-site and via distributed cloud services -- definitions of HR analytics became more precise. This evolution expanded the role of data, analytics, and evidence-based decision-making within HRM \parencite{Kim.2021, Coron.2022}.  

Over the past 25 years, HRM has become increasingly focused on metrics and analytics, accompanied by a substantial growth in data volumes \parencite{Poba-Nzaou.2022}. Today, HRM employs quantifiable metrics to establish key performance indicators (KPIs), enabling the setting and assessment of measurable goals \parencite{Coron.2024}. More broadly, \textcite{Bondarouk.2016} describe how rapid technological advances continue to reshape HRM practices by cultivating a smart, digital environment that enhances the quality of HRM data. Over the past 15 years, digital transformation has enabled stronger HRM ownership among diverse stakeholders, including both HR professionals and organisational members not traditionally engaged in HR tasks. According to \textcite{Bondarouk.2016}, technology enhances data management and decision-making processes, thereby contributing to more effective HRM practices. This trend highlights HR’s evolving role within increasingly dynamic and uncertain work and business environments. 

As noted by \textcite{Minbaeva.2021}, current \enquote{mega trends,} such as digitalisation and artificial intelligence, are fundamentally reshaping the development of HR. In response, HR must define its role within this technological shift and demonstrate clear value to business processes \parencite{Minbaeva.2021}. The movement towards quantification introduces metrics that substantiate HRM's value in terms that align with top management's preference for data-driven language and decision-making \parencite{Diefenhardt.2024}. 

Recently, AI-driven HRM and HR analytics have actively sought to use the technological potential of machine learning, big data, and the generative capabilities of artificial intelligence. AI-based analyses, predictive models, and prescriptive measures utilise available metrics and data to provide actionable insights into complex business processes, especially in situations characterised by increasing uncertainty \parencite{Mer.2023}.  

Although these technologies are often promoted with optimistic narratives of inevitable progress -- or conversely, decline -- their adoption remains a subject of ongoing debate. Proponents contend that AI-based tools can help reduce human biases and enable \enquote{algorithmic inclusion} \parencite[696]{Kelan.2024} of groups that are otherwise marginalised \parencite{Drage.2022}, while also contributing to improved worker well-being \parencite{Deng.2024}. However, the use of data-driven, automated decision-making regarding individuals, their behaviour, or work has sparked considerable discussion about the ethical and moral implications of AI in HRM \parencite{Simbeck.2019, Ayling.2022}. 
Moreover, even though discriminatory effects can sometimes be more readily identified and addressed in algorithms than in human decision-making, concerns about bias persist as a central issue within the evolving and cautious discourse on HR and people analytics \parencite{Kochling.2021, Kochling.2024}.

\subsection{Institutional and organisational structure of AI application in HRM}
Institutional and organisational factors critically shape the introduction of AI into HRM. Recent research underscores the interplay between external policies, internal capacities and the readiness of HR professionals to implement advanced algorithms and analytics. Multiple studies highlight how external policy imperatives -- such as regulatory support, EU-level guidelines and competitive market pressures -- interact with organisational structures and capabilities to influence AI adoption. These works reveal that strong institutional frameworks, leadership commitment, workforce competencies and robust technological infrastructures collectively determine the effectiveness of AI integration in HRM. Organisations that articulate clear strategic objectives, foster collaboration and invest in both human capital and technological resources are particularly successful in leveraging AI for productive outcomes. 

Institutional regulatory policies significantly shape technological change within organisations by addressing sustainability, fairness, risk control and data privacy concerns \parencite{Ogbeibu.2024, EuropeanCommission.2021, EuropeanParliament.2016}. The discourse on AI has intensified in recent years, leading to targeted strategies that facilitate technology adoption by companies \parencite{AfMalmborg.2023}. Malmborg and Trondal demonstrate that national AI policymakers are both constrained and enabled by existing organisational capacities, highlighting the complex relationship between institutional environments and technological adoption. While such frameworks can provide critical momentum for AI integration, successful implementation in HRM ultimately depends on an organisation’s internal ability to leverage these directives. Accordingly, several studies underscore the pivotal roles of organisational culture, leadership support and technical infrastructure in fostering AI readiness \parencite{Bechter.2022, Pan.2022}. 

In addition to institutional settings, organisations play a decisive role in determining if and how they adopt and utilise technologies such as AI tools. Previous digitalisation processes have demonstrated that organisational capacity and readiness significantly influence both the nature and outcomes of transformation initiatives -- for example, in the case of e-HRM systems \parencite{Panayotopoulou.2010}. Firms adopted initial digitalisation and datafication efforts at varying rates. Although personal computers and the Internet are relatively easy to implement across organisations of all sizes, deploying comprehensive e-HRM platforms and advanced data-processing solutions is more complex and typically viable only for firms of a certain scale -- those with multiple branches and dedicated HR departments \parencite{Strohmeier.2009}. Early studies on AI-based tools and evidence-based HR analytics noted surprisingly low adoption rates and limited evidence of actual use and effectiveness \parencite{Marler.2017}.

Recent research confirms similar dynamics for AI, AI-based tools, and People Analytics. Studies consistently show that organisational size, structural centralisation and sector context shape the adoption and effectiveness of AI-based HR technologies across European corporations. Larger organisations with ample financial and technological resources -- as well as a culture of innovativeness, proactiveness and risk tolerance -- particularly tech-focused firms and multinational corporations, embrace AI-driven HR tools more readily \parencite{Baldegger.2020, Fernandez.2021, Lodra.2024, Neumann.2024}. In the ICT sector, strong technological readiness and competitive pressure further drive adoption despite ongoing security and privacy concerns. By contrast, small and medium-sized enterprises (SMEs) and public organisations exhibit lower uptake due to limited resources, bureaucratic hurdles, complex integration challenges and a general lack of AI expertise and awareness \parencite{Medaglia.2022}. Moreover, the potential use cases for AI tools in these settings are constrained by smaller employee bases, fewer application scenarios and limited data availability \parencite{Dutta.2024}.

Organisational leadership and culture consistently rank among the strongest internal drivers of AI adoption in HRM. Pillai et al. \parencite{Pillai.2020} report that top management support critically enhances economic effectiveness, HR readiness and the practical utility of AI technologies. Competition and vendor support further facilitate adoption, whereas concerns about data security, privacy and process-change risks hinder it \parencite{Pillai.2020, Lodra.2024}. Strong work organisation and centralised HR roles also correlate with smoother AI integration. \textcite{Fernandez.2021} finds that an HR department’s technical expertise, strategic positioning and storytelling capabilities are key to successful HR analytics adoption. Executive champions secure necessary budgets, reduce internal resistance and guide the integration of new digital tools \parencite{VanNoordt.2022}. Meanwhile, HR professionals need trust and competence to blend AI insights with human judgement, ensuring alignment with ethical, fairness and transparency standards. Nonetheless, many organisations have yet to realise AI’s full benefits \parencite{Chowdhury.2023}, often due to social and cultural barriers that demand targeted leadership training and qualification. Skill gaps in analytics and data management remain a recurring obstacle for HR managers \parencite{Baldegger.2020}.

Technical and infrastructural readiness is another critical dimension \parencite{Lodra.2024}. Organisations frequently face challenges such as inconsistent HR-metrics standards, fragmented data and incompatible systems \parencite{Fernandez.2021}. Effective AI deployment depends on robust data management, seamless platform integration and well-defined performance metrics to generate reliable insights. When organisational data are poor in quality or dispersed across silos, the credibility of AI-driven recommendations declines, undermining user confidence and broader acceptance.  \textcite{Neumann.2024} further highlights that an organisation’s AI maturity level -- from isolated pilot projects to full-scale operational infrastructures -- significantly shapes its adoption trajectory.

Beyond internal capabilities, bureaucratic regulations shape innovation trajectories. Research indicates that administrative \enquote{red tape} can facilitate diffusion, although its influence varies \parencite{Revillod.2024}. Moderate formalisation and accountability frameworks help ensure that AI adoption remains transparent and aligned with organisational goals. However, excessive red tape can hinder timely experimentation, while insufficient oversight may give rise to ethical or fairness concerns -- especially in human-centred functions such as HR \parencite{Revillod.2024}.

Finally, studies emphasise the importance of collaboration and stakeholder engagement \parencite{VanNoordt.2022}. Cross-departmental teamwork, partnerships with technology vendors and the early inclusion of employee representatives foster solutions that address practical needs \parencite{Lodra.2024, Kalff.2022b}. Such participatory approaches also build trust and clarify user roles within new AI-driven workflows.

\subsection{AI in HRM: Effects on work tasks and managerial identity}\label{subsec:worktasks}
New AI-based use cases and tools have entered the HRM market in the past decade. Industrial discourse often conveys optimism about these technologies’ disruptive potential for enhancing efficiency and productivity. This enthusiasm echoes the capabilities attributed to AI -- though the extent of these capabilities remains under scrutiny \parencite{Narayanan.2024}. Various studies highlight AI’s contributions to HR, including candidate ranking, churn-risk analysis, workforce management, CV parsing and highly individualised training and development programs (cf. Table~\ref{tab:aitools} for an overview of HRM tasks across the employee life cycle that are supported or taken over by AI functions). By integrating company data -- often in real time -- into pre-trained AI models, organisations can bolster strategic capacities, support informed decision-making and accelerate processes. Literature reviews and case studies document this ongoing debate on AI-driven HRM, emphasising process optimisation and efficiency gains \parencite{William.2023, Muridzi.2024}. While AI-based HR and HR analytics thrive in centralised corporate structures, decision-making authority is increasingly shifting to operational teams and middle management, where AI tools offer guidance and support \parencite{Lehmann.2024}.  

\begin{table}
	\centering
	\caption{Studies on AI tools for different tasks in the \enquote{Employee Life Cycle}.}
	\label{tab:aitools}
	\begin{tabularx}{\textwidth}{cXY}
		\toprule
		\textbf{HR Life Cycle} & \textbf{Tasks} & \textbf{Sources} \\
		\toprule
		Attraction & Creating, formulating and optimising job ads; posts on professional networks; informative Chat-Bots; vacancy prediction; automated sourcing & \cite{Rukadikar.2023}, \cite{Tsiskaridze.2023}, \cite{Zhai.2024}\\
		Recruiting & CV Parsing, matching, and ranking; initial interaction; generating interview questions; video analysis of assessments; automated documentation; psychological profiling & \cite{Budhwar.2022, Budhwar.2023}, \cite{Chowdhury.2023}, \cite{Jatoba.2023}, \cite{Malik.2023}, \cite{Revillod.2024}, \cite{Tsiskaridze.2023}, \cite{Zhai.2024}\\
		Onboarding & Structured onboarding guide; on-boarding chat-bots; individual tailored onboarding; FAQ handling; process automation  & \cite{Chowdhury.2023}, \cite{Malik.2023}, \cite{Mer.2023}, \cite{Zhai.2024}\\
		Retention & Measure and predict motivation, sentiment or performance; targeted motivational strategies; churn and flight risk analysis;  & \cite{Chowdhury.2023}, \cite{Coron.2024}, \cite{Mer.2023}, \cite{Zhai.2024}\\
		Development & Tailored qualification road map; succession management; substitution management & \cite{Arora.2021}, \cite{Budhwar.2022, Budhwar.2023}, \cite{Coron.2024} \cite{Jatoba.2023}, \cite{Malik.2023}, \cite{Zhai.2024}\\
		Management & Performance appraisal; Feedback generation; targeted compensation management; goal setting and tracking	& \cite{Budhwar.2022, Budhwar.2023}, \cite{Chowdhury.2023} \cite{Coron.2024}, \cite{Malik.2023}, \cite{Mer.2023}, \cite{Zhai.2024}\\
		Offboarding & Feedback generation; Certificate creation; Exit trend analysis; automated offboarding processes & \cite{Charlwood.2022}, \cite{Zhai.2024}\\
		\bottomrule
	\end{tabularx}
\end{table}
\textcite{Arora.2021} demonstrate that AI-backed people analytics significantly enhance HR’s efficiency and productivity. Consequently, the focus of AI in HRM has shifted from routine administrative tasks to complex processes such as talent acquisition, training and development, employee retention, engagement, and performance appraisal. \textcite{Tsiskaridze.2023} emphasise AI’s dual nature: it promises efficiency gains but raises ethical and legal concerns related to fairness and data protection. Effective human–machine collaboration thus requires careful design to mitigate perceptions of unfairness. \textcite{Mer.2023} further reports that AI tools can reduce costs, decrease attrition and boost productivity. Examples include natural language processing (NLP) and robotics streamlining HR self-service tasks, and advanced analytics bolstering DE\&I initiatives.

Critical scholarship warns that AI may degrade work quality by intensifying surveillance, restricting autonomy or displacing employees. \textcite{Charlwood.2022} argue that HR professionals play a crucial role in guiding AI integration, balancing its benefits and drawbacks to ensure people-centred applications. This tension reshapes managerial roles, responsibilities and identities within HR. \textcite{Lippert.2024} identifies which managerial roles AI might replace, which may evolve into human–AI collaborations and which will retain or gain importance. Routine monitoring tasks traditionally handled by middle managers may shift to AI, while new \enquote{managerial meta-roles} \parencite[10]{Lippert.2024} emerge to oversee AI-related functions in employee management. \textcite{Malik.2023} propose a strategic framework for HR managers implementing AI-assisted HRM, stressing the need to balance AI capabilities with human factors by fostering continuous learning, ensuring ethical AI use and maintaining transparent communication with employees about AI’s role in HR processes.

A substantial number of AI tools aid decision-making by offering additional data, big-data analytics, predictions and action-oriented prescriptions. However,  \textcite{Bader.2019} argue that algorithmic decision-making can spatially, temporally, rationally and cognitively distance stakeholders from their own decisions, even as they remain affected by them. These mismatches give rise to three coping strategies: deferring decisions, employing workarounds and manipulating data, all of which impact professional identities. \textcite{Loscher.2022} identify another challenge to HR’s professional identity: HR managers and data analysts operate with divergent understandings of HRM practices and professionalism. Data analytics introduces quantifiable insights into a domain traditionally guided by experience, interpersonal knowledge and intuitive judgment. Consequently, tensions emerge between data analysts -- who drive AI in HRM -- and traditional HR managers. Yet \textcite{Loscher.2022} demonstrate how traditional HR can leverage these insights to strengthen its influence on top management. \textcite{Diefenhardt.2024} maintain that a growing data orientation helps HRM achieve \enquote{strategic recognition} within corporate governance to enhance its influence. Securing this recognition, however, requires micropolitical tactics to establish the credibility of HR analytics. These challenges mirror those experienced during e-HRM implementation, including administrative inertia, team fragmentation, local resistance and ambiguous identity \parencite{Barrett.2013}.

Several authors argue that, for a genuine transformation via e-HRM and AI, HR must move beyond process automation to embrace deeper shifts in its self-concept, innovative capacity and service logic, thereby positioning itself as a strategic partner in organisational innovation \parencite{Barrett.2013, Lehmann.2024, Jatoba.2023}. HR analytics contributes to accountability in people management by enhancing the visibility and transparency of HR practices and outcomes. Moreover, it can guide the design of HR processes that embed accountability, ensuring alignment with strategic goals and continuous evaluation against performance metrics. By connecting HR metrics with those of other organisational units, HR analytics provides a comprehensive view of how HR activities interact across the organisation \parencite{Loscher.2023}.

\subsection{The ethical implications of AI in HRM}\label{subsec:ethics}
Various studies critically evaluate AI applications in HRM, examining their ethical and legal implications. Scholars emphasise AI’s ambivalent nature and potential risks, including ethical dilemmas, data privacy concerns, the need for transparent decision processes and the principle that final decisions should rest with humans \parencite{Malik.2023, Charlwood.2022, Tsiskaridze.2023}. A significant strand of this research focuses on AI’s role in Diversity, Equity \& Inclusion (DE\&I). AI is not uniformly regarded as a threat; instead, it can serve as a tool to enhance fairness through advanced analytics and foster inclusion \parencite[696]{Kelan.2024}. Moreover, discriminatory patterns are often easier to detect and correct algorithmically than in human practice, a point central to the cautious discourse on HR and People Analytics \parencite{Kochling.2021, Kochling.2024}.

By contrast, sceptical perspectives warn against inflated expectations of AI. Critics highlight concerns about data security, informational self-determination, heightened surveillance, intensified worker control and new forms of discrimination arising from biased data, models or assumptions -- compounding existing human biases. They also challenge claims of reliable prediction in HR analytics, dismissing such AI “prediction machines” as “elaborate random number generators” \parencite[25]{Narayanan.2024}. Consequently, the true capabilities and limitations of HR and People Analytics remain a subject of active debate \parencite{McCartney.2022}.

Although AI’s overall effectiveness is sometimes questioned, critical studies focus on its practical applications. Challenging optimistic narratives, \textcite{Schellmann.2024} documented discriminatory risks in AI-based recruitment and hiring tools: marginalised groups were excluded due to arbitrary and biased training data. On a more technical level, researchers highlight hazards such as biased datasets, insufficient control mechanisms and opaque model outputs \parencite{Ayling.2022, Kochling.2021, Fabris.2024}. Poorly trained AI systems -- with nontransparent weights, arbitrary selection criteria or undisclosed data transformations -- undermine efforts to create diversity- and equity-aware workplaces \parencite{Du.2024}.

In a technical evaluation of an algorithmic video-analysis tool for recruiting,  \textcite{Kochling.2021} found that the underrepresentation of specific gender and ethnic groups in the 10\,000-video training set led to biased recommendations. The tool overestimated the suitability of underrepresented candidates for interviews, thereby perpetuating and sometimes intensifying existing inequalities. Beyond discrimination, privacy concerns arise when employee data become more transparent while their origins, transformations and evaluative criteria remain opaque \parencite{Simbeck.2019}.

Moreover, labour politics highlight the disciplining and controlling effects of data on employee behaviour, facilitating new forms of objectification, work surveillance and the rationalisation of organisational processes \parencite{Weiskopf.2023, Williams.2024}. In HRM, these fine-grained insights -- on behavioural data, psychological profiles, social media activity and work processes -- enable detailed profiling, categorisation and decision-making regarding individual candidates or employees. \textcite{Charlwood.2022} therefore urge HR professionals to engage proactively in AI oversight to ensure fairness, promote the ethical handling of employee data and preserve work quality.

%A key priority remains establishing robust processes to address and mitigate risks stemming from AI technologies in HR. Given the high-stakes nature of HRM applications -- which directly affect employees’ career prospects -- EU legislation classifies such AI tools as \enquote{high risk} and mandates comprehensive audits \parencite{EuropeanCommission.2021}. In response, \cite{Mihaljevic.2023} advocate a contextual, participative auditing approach: audits should be integrated into existing workflows, rely on real (not synthetic) data and incorporate diverse perspectives -- from advocacy groups and researchers to vendors and employees. While using real-life data strengthens AI assessment, it also requires rigorous data-security safeguards for the individuals represented. Because HR functions both as a key stakeholder in the AI toolchain and as employees’ representative, these participative audit practices will directly involve HR teams.

\section{Research Design}\label{sec:methods}
The empirical data were collected as part of the research project \textit{TranKI -- Standards for transparent AI}\footnote{Funding provided by Hans Böckler Foundation. Grant number: 2022-797-2. Project duration: 10/2023–09/2026. Further insights: \textcite{Kalff.2025c}, project dataset: \textcite{Simbeck.2025}.}
 to capture current trends, use cases and challenges of AI systems in HR management. To address our research questions, we conducted semi-structured interviews, focus-group discussions and a survey to assess how HR managers perceive AI adoption and AI-based HR tools within their organisations and professional contexts. We employed a mixed-methods design -- drawing on multiple data sources to cross-validate findings and integrate qualitative and quantitative approaches \parencite[611–644]{Bryman.2011}.  

First, our research design comprised qualitative interviews with four stakeholder groups: a) experts in AI and AI-based HR tools; b) developers and distributors of these tools; c) HR managers working with AI-augmented systems; and d) representatives from civil-society NGOs (data, privacy and diversity advocates) interested in AI’s dynamics in HR. Participants were purposively selected based on their expertise \parencite{Glaser.2009} in one or more of the following areas: software development; practical implementation and challenges; comprehensive knowledge of HRM and technological innovation. We also documented the societal relevance of AI systems and related ethical and moral considerations. The interview topics included: AI use in HR management; changes in HR roles and workflows; transparency requirements; and technical and organisational implementation. Each interview lasted 45–60 minutes. All sessions were transcribed and analysed using MaxQDA \parencite{Radiker.2020}. Table~\ref{tab:interviews} summarises the interview sample. For analysis, we applied interpretative content analysis \parencite{Drisko.2016, Kuckartz.2023}. Deductive categories were derived from a systematic literature review on topics such as AI in HRM, explainable AI, transparency, fairness and bias. Inductive categories emerged directly from the interview data. This combined approach ensured analytical rigor while preserving openness to participants’ perspectives and insights beyond the original research scope \parencite[76–77]{Kuckartz.2023}.  

\begin{table}
	\centering
	\caption{Overview of qualitative expert interviews}
	\label{tab:interviews}	
	\begin{tabularx}{\textwidth}{ccXl}
		\toprule
		\textbf{No.}	&	\textbf{ID}	&	\textbf{Type of institution}	&	\textbf{Person and field of expertise} \\
		\toprule
		1	&	ASSOC1	&	Professional association HR	&	Head of Strategic HR\\
		2	&	COD1	&	Consultancy for employee representation	&	IT-Consultant\\
		3	&	COM1	&	Software company	&	Head of HR\\
		4	&	COM2	&	Telecommunications company	&	Head of Diversity, Equity and Inclusion\\
		5	&	COM3	&	Digital Services company	&	Head of Recruiting\\
		6	&	COM4	&	Pharmaceutical company	&	Head of Talent Management\\
		7	&	COM5	&	Start-up DE\&I services	&	Founder; CEO\\
		8	&	COM6	&	Mechanical engineering company	&	CHRO\\
		9	&	DEV1	&	Software development company DE\&I tool	&	CTO\\
		10	&	DEV2	&	Software development company talent management	&	Head of Consulting\\
		11	&	DEV3	&	Software development company workforce management	&	Head of technical development\\
		12	&	NGO1	&	Consumer protection	&	Researcher\\
		13	&	NGO2	&	NGO for ethics in tech	&	Activist\\
		14	&	NGO3	&	AI Strategy, DE\&I	&	AI Strategist, Science Communicator\\
		\bottomrule
	\end{tabularx}
	
\end{table}

%Second, we conducted three group discussions with advisors and representatives of works and staff councils. In Germany, the advisors at the staff units of the works and staff councils are generally not direct employees of the respective companies. They can be hired by works and staff councils to provide advise on complex issues with explicit expertise to support co-determination. They have a deeper level of knowledge, especially regarding technological topics such as artificial intelligence. In this setting, it was possible to acquire a more fundamental understanding of the shared life-world of institutional co-determination concerned with shaping AI in the interests of employees \parencite{Thaa.2020, Bohnsack.2010b}.

Second, we conducted three group discussions with advisors and representatives of works and staff councils. In Germany, these advisors are typically external specialists hired by councils to support co-determination on complex issues. They bring deep expertise -- especially on technological topics such as artificial intelligence -- and thus provide insight into the institutional life-world of co-determination in shaping AI for employees’ interests \parencite{Bohnsack.2010b}.  
The group discussions focused on three questions: 1) Which tools are applied at the companies? 2) How could co-determination participate on technology or tool selection and implementation processes? 3) What types of transparency need factory and staff councils to make informed decisions whether or not a tool is green lit for HRM to use? This aspect is crucial in German labour rights since any technology that can measure, surveil or control employee performance or behaviour must be subject to employee co-determination (Works Constitution Act -- BetrVG §87, (1) 6).\footnote{The English version of the law can be accessed on the official federal website: \url{https://www.gesetze-im-internet.de/englisch_betrvg/englisch_betrvg.html\#p0503}.} The development of AI tools, their implementation, and finally their actual application are the responsibility of different groups of people, organisational entities or even entire organisations in the AI value chain. This diffusion requires \enquote{ethics work} \parencite[3]{Widder.2023} to overcome the ethical dilemmata stretching to different levels. Co-determination does take on the role of controlling and disputing AI technology in the interest of the workforce and employees. Group discussions with council advisors thus enriched our understanding of which technologies are deployed, how they support different HR tasks, and how political negotiations shape their introduction.

Third, we administered a survey to 427 HR managers in Germany, gathering data on attitudes toward AI and AI literacy; organisational background; which HR activities are already supported by AI; and what motivates employers to adopt AI technologies. After removing implausible entries, 410 valid responses remained. The survey was fielded in February 2025 by an ISO 20252:2019-certified sampling provider and pre-tested with 15 participants in December 2024. Data were analysed in R using standard packages. Our analysis focuses on descriptive summaries of the survey findings to answer our research questions.

\section{Results}\label{sec:results}
\subsection{AI's transformative effects on HRM}\label{subsec:aiapplications}
\begin{figure}[tb]
	\includegraphics[width=\textwidth]{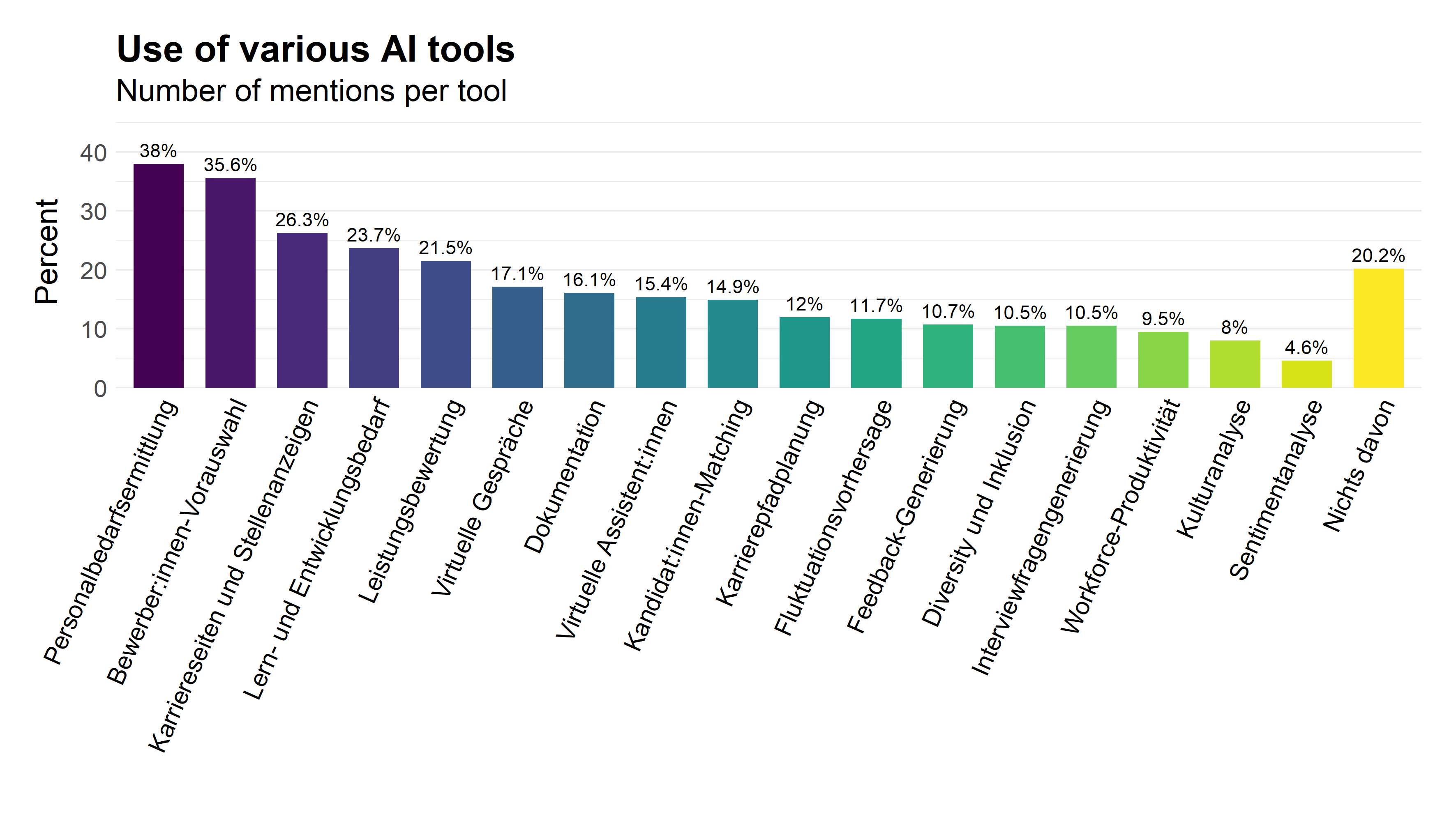}
	\caption{Overview of different tools that are utilised by German HR managers (source: own data)
		\label{fig:tools}}
\end{figure}

AI-based HRM technologies are diverse, yet their application within organisations remains limited to specific use cases. In our survey, the most commonly referenced HR activities employing AI were workforce planning, (semi-)automated candidate screening, creation of career pages and job advertisements, assessment of training and development needs, and performance evaluation. As illustrated in figure~\ref{fig:tools}, a notable share of respondents (over 20\,\%) reported that their departments do not use any AI tools. This variation indicates that, despite AI’s potential to enhance operational efficiency, its adoption in HR is uneven across organisations, reflecting differences in internal priorities, digital infrastructure, organisational readiness and risk perceptions.

A further nuance in our data concerns the types of AI solutions applied in HR processes. While traditional machine-learning applications for predictive and prescriptive HR analytics maintain a presence, respondents generally perceived their use as limited. In contrast, large language models (LLMs) and other generative AI tools have recently gained visibility and relevance in HR. Since OpenAI released ChatGPT in November 2022, industry exhibitions and public demonstrations have showcased how these models can synthesise documents, finalise contracts and automatically generate job postings from user prompts. Such practical use cases underscore LLMs’ rapid rise in prominence: their accessibility lowers technical barriers and fosters both formal and informal experimentation.

Our research indicates that even in companies where official policy restricts or prohibits generative AI, employees regularly adopt these tools on personal devices -- of the 410 valid survey responses, 183 reported informal use of AI tools regardless of employer policies or the availability of company-provided systems. By contrast, AI solutions requiring complex, company-specific data cannot be used informally, as they are accessible only through an official rollout.  

Chatbots are among the simplest AI applications in HRM, automating repetitive queries and basic administrative tasks to free up HR staff resources (COM1). Although well established in customer-service contexts \parencite{Ciesla.2024}, chatbots have increasingly been adopted in HR services, delivering rapid data retrieval (e.g., reminders about company policies or benefits), standardised Q\&A workflows and preliminary screening of external candidates. An interviewee from a large software company (COM1) -- who both develops and uses AI tools for HRM -- explained how a chatbot now handles various routine tasks in their HR department. Consequently, employees who previously performed these duties were laid off:  

\blockquote[COM1]{And that used to be done by our clerks, with lots of lists from different systems and tools, Excel sheets and merging here and there. \textelp{} This is now tool-supported; it runs through our system. And have the jobs disappeared? Yes, they have disappeared. Do I regret that? I don't think so; I mean, the idea is, of course, that people do completely different tasks, yes, because they are qualified accordingly. Is that an issue in the future? Yes, it's also an issue for many professions and job roles. And I also think it's important that it's discussed in public.}
A second application area involves AI-enhanced recruitment. Although automation in recruitment is not new, recent advances have emphasized voice-to-text transcription, generative summaries and algorithmically guided decision-making. One interviewee at COM3 described integrating AI-generated transcripts and automated interview summaries into their hiring process. This technology reduces extensive note-taking, allowing interviewers to focus on real-time, face-to-face engagement. Moreover, AI-generated summaries and candidate scorecards streamline decision-making when evaluating large applicant pools. As one interviewee explained, \enquote{Nobody likes going into recruiting to spend ages formulating scorecards or formulating our LinkedIn posts \textelp{}. They go into recruiting because they are good with people and cherish interaction with people} (COM3). Under this system, routine tasks are delegated to machines, freeing HR staff to concentrate on more engaging, interpersonal work.

Another key family of HR AI applications concerns talent management. While employee learning, development and succession planning have long been central to HR’s strategic role, AI-based tools promise more granular, data-driven forecasting. Interviewees DEV2 and COM4 highlighted solutions using machine-learning algorithms to detect skill gaps, predict career trajectories and recommend individualized upskilling or mentoring programs. The logic is that organisations can no longer rely solely on external hiring to address skill shortages; developing internal talent has become a strategic priority.

\blockquote[COM4]{But when I look at AI in the context of a professional career, it has always been about the pivotal point of talent management, from talent acquisition to talent retention, talent development, talent identification, and the matching of opportunities. These are not just vacancies but also projects, learning nuggets, learning content, further development options, and mentor-mentee relationships. That's what AI has always been about for us: predicting and forecasting the development of an individual’s career in whatever form it may take.}
While recruitment primarily targets external labour markets, AI-supported internal talent management relies on extensive employee data repositories -- covering current and potential roles and aligning with strategic corporate planning. These systems use historical metrics to forecast career trajectories. For example, employees with strong performance scores and demonstrated leadership potential may receive tailored development modules or short-term project assignments. However, despite their promise of personalised career management, AI integration often stalls due to fragmented data governance and a lack of centralized, reliable infrastructures (COD1; COM6).

Another AI use case emerging from our interviews is resource scheduling and workforce management. DEV3, a provider of shift-planning software for nearly two decades, evolved its stochastic tool by incorporating machine learning to predict staffing demand. By analysing past demand peaks, seasonal patterns and local events, the solution generates schedules that balance employee preferences and required qualifications (e.g., first-aid certification or specific machine training) with operational needs and legal constraints. The DEV3 interviewee noted that, although the algorithms could be labeled \enquote{AI,} the term also serves as a powerful marketing hook. In practice, AI branding attracts clients seeking to modernize or future-proof their workforce processes.  

Nevertheless, a gap often emerges between AI’s perceived benefits and its practical feasibility across organisational contexts. Our group discussions and interviews confirmed that many SMEs fail to see immediate returns from complex AI-driven applications, which typically depend on large volumes of centralised data and stable, standardised processes. Without these prerequisites, advanced AI use cases in HR are seldom economically or logistically viable. By contrast, generative AI tools are an exception: their low adoption threshold enables individual employees or small teams to integrate them seamlessly into everyday workflows -- even unofficially.

Moreover, legal and regulatory constraints can further complicate AI deployment. The European AI Act classifies certain HR applications (e.g., candidate screening or predictive performance management) as high-risk under Article 35 of the EU AI Act, thereby triggering stringent auditing requirements \parencite{EuropeanCommission.2021}. According to our interviewees, some organisations adopt simpler chatbots or generative-text assistants to sidestep these compliance obligations, shaping their HR rollout accordingly. For example, DEV3 intentionally limits its forecasting software -- used to predict patient inflows or store visits -- so that it does not process personal data and thus avoids the more onerous data-protection and co-determination requirements.

In summary, HR AI tools fall into two broad categories: 
(1) generative models and chatbots that require minimal specialised datasets and automate primarily text- or communication-based tasks; and 
(2) predictive-analytics systems trained on proprietary internal data to guide decisions in recruitment, succession planning or workforce scheduling.  
These applications underscore AI’s growing sophistication in HR, yet deployments remain largely confined to core functions. While some organisations leverage AI to optimise and scale recruitment, talent development and workforce allocation, others remain cautious or restrict themselves to minimal use. This variation reflects organisational context: data quality, firm size, digital maturity and managerial support collectively shape the feasibility and scope of AI-driven HR initiatives. 

In the following section, we present our findings on the organisational structures and cultural attitudes crucial for AI adoption, and how these technologies, once integrated, transform the nature of HR work.

\subsection{Organisational Aspects of AI Adoption}\label{subsec:orgaspects}
Many of the strengths and limitations associated with using AI in HRM arise from broader organisational structures and processes. Throughout our interviews and group discussions, participants consistently emphasized that factors such as corporate culture, internal governance, co-determination practices, and data management collectively facilitate or impede the implementation of AI-driven HR tools.

A key factor influencing AI adoption is how AI is perceived and defined within organisations. For example, group discussions often devolved into debates over the meaning of AI. Many participants’ perceptions were strongly shaped by their exposure to generative AI applications popular in the media or in personal use (e.g., ChatGPT). As a result, some participants did not recognize machine-learning systems that underpin traditional HR analytics -- such as those used to measure turnover risk or identify workforce trends -- as \enquote{actual AI.} The flexible and often ambiguous use of the AI label can therefore become strategically significant. Vendors may leverage AI branding to generate interest and support business cases. Conversely, the AI aspects of HR tools may be downplayed to avoid scrutiny from co-determination bodies.

From a legal perspective, risk classifications under the AI Act provide an important benchmark for evaluating AI technologies and their potential functionalities. As DEV3 described: \enquote{We're so far away, so to speak, that we don't even have personal data, so we don't have a GDPR or DSGVO thing and we don't have an AI Act thing either, because the technologies we use are relatively discreet, and the data we process is not critical. It's just aggregated data, there's not even a single incident where a person plays a role} (DEV3).

Additionally, Germany’s regulatory tradition, exemplified by the Works Constitution Act, supports a strong system of worker representation. For instance, when an AI system is used to evaluate employee performance or select candidates for promotion, it typically triggers co-determination rights. Employee representatives, aware of the risks of algorithmic discrimination or opaque \enquote{black box} systems, frequently demand strict oversight. As noted by the consultant for co-determination bodies (COD1), this often includes advocating for contractual blocking clauses.

In practice, requirements for works council approval of AI use in HR can help ensure that AI projects protect employee interests. However, globally active companies sometimes outsource HR functions to affiliates or global service centres. By doing so, these functions are removed from the jurisdiction of local works councils and the scope of strict EU or German legal requirements. According to COD1, such externalisation strategies are increasingly common when organisations wish to avoid negotiations regarding sensitive AI-based analytics.

Further organisational challenges arise in the way data are collected and managed. HR analytics and AI systems typically require large, high-quality, and consistently structured datasets to be effective. Many German companies -- especially in the SME sector -- lack centralised digital infrastructures or have historically used independent HR systems at each regional office or subsidiary. As COM1 reported, these fragmented data landscapes impede coherent analysis and limit the potential of advanced analytics:  

\blockquote[COM1]{What has always accompanied me in my life at this company, or us in HR, has been the topic of standardisation and automation. So, two things come together here. On the one hand, we are an American company, where the way of thinking is quite different \textins{from companies of German origin}. And then there's the fact that we're a technology group. This means there has always been an interest in trying out, using and further trying out technology developed within the company or bought.}
When organisations maintain decentralised or non-standardised data practices, opportunities for advanced AI applications are significantly reduced. Additionally, the attitudes of key decision-makers towards digital transformation play a critical role. Companies with a strong technology-oriented culture, particularly those in software, IT, or globally focused industries, are generally more open to systematic standardisation. In contrast, traditional or regionally based companies often favour customised, local solutions, allowing business processes to be managed at the departmental level. Although this autonomy provides flexibility, it can also lead to fragmented data resources and hinder the consistent implementation of AI.

\begin{figure}[t]
	\includegraphics[width=\textwidth]{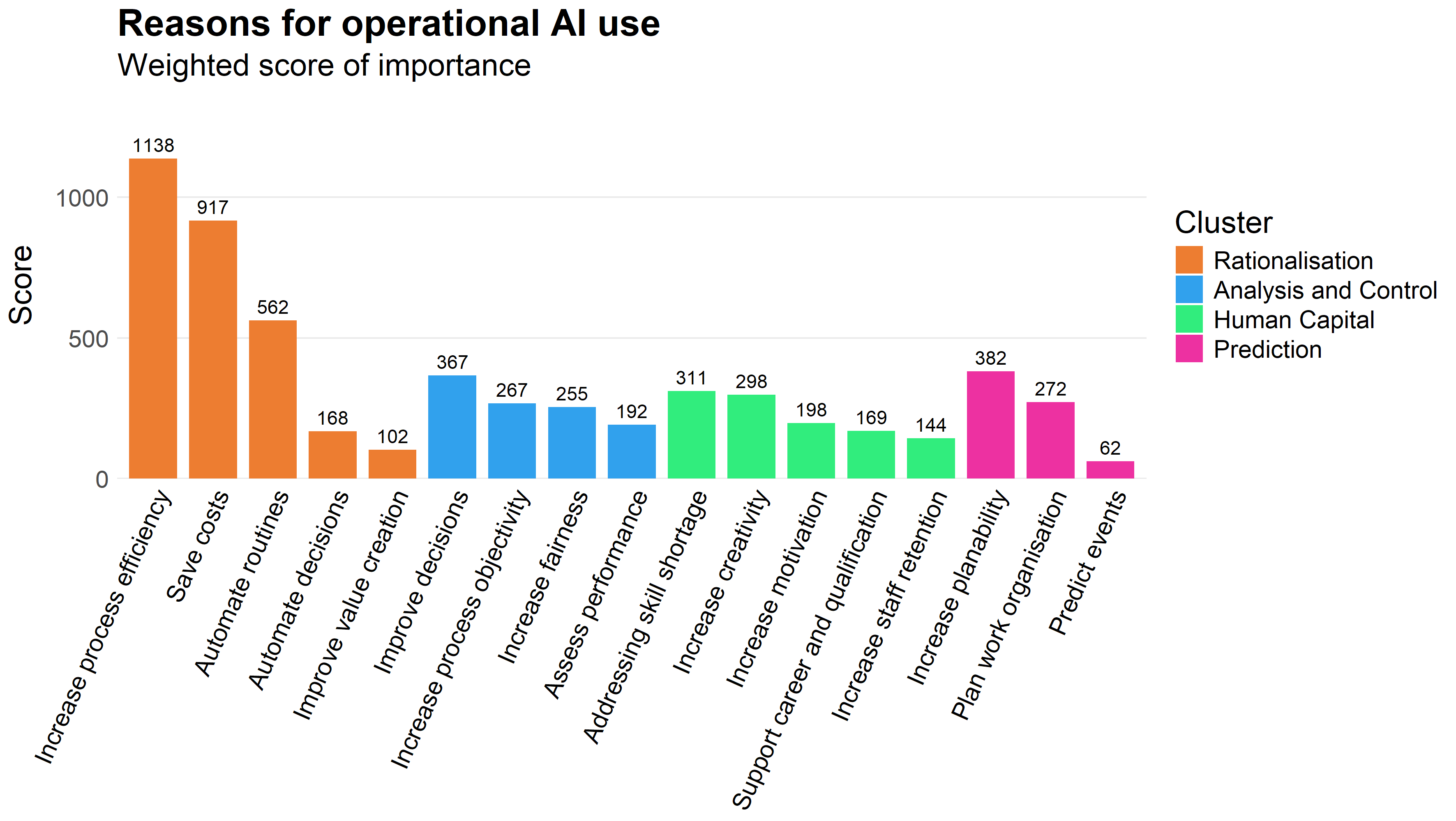}
	\caption{Overview of strategic goals behind AI use in German HR departments. We asked HR managers to rank their organisations' five most important reasons for implementing AI tools. A first-place ranking was assigned 5 points, while a fifth-place ranking received 1 point. Scores are cumulated. (source: own data)}
	\label{fig:reasons}
\end{figure}

Persistent skill shortages in Germany also influence AI adoption from an organisational perspective. Tools designed to manage administrative burdens -- such as screening large numbers of applicants -- are particularly valuable when application volumes are high. However, complex AI-driven screening systems may inadvertently exclude suitable candidates in sectors already struggling to find qualified applicants. This mismatch helps explain why some organisations are reluctant to invest in AI-based ranking or filtering processes: they are concerned about missing out on scarce talent or see limited economic justification for such investments. As a result, many organisations opt for minimal digital enhancements to reduce administrative workload, rather than undertaking a comprehensive transformation of their recruitment processes.

Beyond data and organisational culture, the strategic rationale for AI investments is another important factor. Our survey results (figure~\ref{fig:reasons}) show that when asked about their employers' rationale for introducing AI, most respondents view AI primarily as a means to increase efficiency, reduce costs, or streamline existing HR processes. In these cases, the \enquote{top-down} motivation for AI adoption (COD1) is focused on rationalisation, rather than on more strategic goals such as workforce development or evidence-based decision-making. Organisational structures driven by cost-efficiency tend to prioritise immediate returns on investment, which can limit support for more exploratory or transformative uses of AI. This context helps explain our finding that predictive or prescriptive applications of HR analytics play only a minor role in our empirical observations.

Interestingly, organisational strategies for implementing AI-based HR analytics and predictive tools often clash with contemporary management styles that prioritise self-organisation and reduced hierarchy. Interview and group discussion insights (COD1, COM4) indicate that tightly integrated HR analytics are frequently seen as overly rigid forms of micro-management, contradicting efforts to foster autonomy and employee empowerment \parencite{Kalff.2020, Kalff.2023}. Predictive workforce analytics may also reintroduce hierarchy quantification into daily practices, undermining stated values of trust and responsiveness.

In summary, the organisational aspects of AI adoption in HR are shaped by the interplay of culture, data centralisation, skill shortage, co-determination norms, and strategic priorities. These structural complexities are significant and should not be underestimated. Multiple corporate management philosophies may even conflict with the intended roles of AI-based HR tools and the organisational foundations required for their effective use. As a result, technological adoption does not happen automatically simply because it promises specific benefits. Instead, AI technologies require organisational structures that both accommodate and support their practical use.

\subsection{AI-augmented HR?}\label{subsec:aiaugmented}
The expansion of AI tools, as described above, raises a fundamental question: how do these systems, in practice, transform the role and profession of HR managers? The majority of research literature identifies two main trends: the automation or the augmentation of human labour in HR. However, some critical voices contend that this distinction is overly simplistic, arguing that automation and augmentation occur simultaneously, interactively, and are mutually dependent \parencite{Raisch.2021}. Evidence from our interviews and group discussions supports this latter perspective. According to the HR managers we interviewed, AI-augmented HR involves automating routine and burdensome tasks of lower perceived value for corporate value creation. This automation enables HR professionals to focus their efforts on more valued activities, such as interacting with people, developing strategies, or fostering talent. Interestingly, our survey results show that corporate strategies for the use of AI to improve value creation are only of very low importance in the eyes of employees (Figure~\ref{fig:reasons}).

We identify three interconnected shifts: First, AI enables HR practitioners to spend less time on repetitive tasks and more on “value-creating” (COM2) activities, such as direct interpersonal interactions and strategic business partnering. Second, the integration of AI into decision-making processes requires HR professionals to develop new competencies, emphasising both technical and analytical skills alongside interpersonal expertise. Third, these developments provoke debates about the nature of human agency and the ethical boundaries of HR practice, as AI reshapes task allocation and redefines professional identity.

Automation plays a central role in reconfiguring HR work. Across various organisational contexts, participants described AI as an opportunity to eliminate or reduce low-value or tedious tasks, such as data entry, scheduling, or manually screening applicant résumés. Companies like COM1 and COM3 present AI-based chatbots and automated transcript tools as ways to free recruiters and HR professionals to focus on meaningful face-to-face interactions with candidates and employees. This shift is generally viewed positively. Delegating repetitive tasks to AI, it is argued, allows HR professionals to concentrate on building personal relationships, developing retention strategies, and planning organisational development. Interviewee  COM2 stated: \blockquote[COM2]{The reason this business case exists in the first place is the hope that it will increase efficiency or take unpleasant repetitive tasks away from employees so that they can devote more time to value-creating tasks. \textelp{} If it doesn't create value, nobody likes doing such mindless things; finding people who want to spend a long time on them is usually very difficult.}
Although many HR managers praise the reduction of administrative duties as a benefit, this transition is not without challenges. Some interviewees noted that these “low-level” tasks historically comprised a significant part of clerical roles in HR; as a result, the introduction of AI has a direct impact on job design, career paths for entry-level HR professionals, and overall staffing needs. Automation can make certain HR functions obsolete or significantly reduce their scope. In response, some companies invest in reskilling or igning staff to roles that require more advanced capabilities. The effects of this shift depend largely on corporate strategy: technology may be used to reduce administrative positions or, alternatively, as an opportunity to reposition employees into more complex or interpersonal roles.

A critical aspect of AI-enhanced HR is its influence on talent development and career pathways. Several respondents (COM4, DEV2) described how AI-based systems support personalised upskilling by matching employees with appropriate learning modules, mentors, or projects. In this model, AI not only automates tasks but also shapes strategic HR processes, providing detailed insights into workforce potential and guiding employees towards targeted professional development. COD1 noted that strong frameworks for employee representation often serve as a counterbalance to individually tailored, algorithmically determined career trajectories, especially when these are enforced as rigid targets. The effectiveness of such systems, however, relies heavily on transparent communication about the underlying logic and objectives.

At the same time, the HR management profession is experiencing a redefinition of its core identity. Traditionally, HR’s unique value has been grounded in expertise in human dynamics: negotiation, coaching, conflict resolution, and empathetic communication. Comprehensive interview data confirm that these “human-centred” skills remain essential, even as AI alters the distribution of tasks (COM2). Organisations with more advanced AI integration often report that automation gives HR a more strategic focus, positioning it as a business partner for employees, technology, and senior management. 

This shift highlights the need for dual competencies in HR. On one side, HR professionals must embrace AI-generated, data-driven insights to advise leadership on workforce strategy. On the other, they must retain a humanist perspective, ensuring that employees are not reduced to data points within an opaque optimisation process (COM5).

Another important aspect of AI-augmented HR is the ethical management of data. The “garbage in, garbage out” principle is widely recognised (DEV2, COM5). If the underlying data reflects historical bias or incomplete employee profiles, AI models risk reproducing or amplifying those biases. HR managers who use these tools for recruitment or career advancement decisions must possess both the expertise and the organisational mandate to question algorithmic recommendations (NGO2, NGO3). Without such critical oversight, the apparent objectivity of AI could legitimise flawed decision-making \parencite{Drage.2022}.

These observations confirm that the introduction of AI in HR not only changes work processes but also transforms the nature of HR professionals’ responsibilities. While certain routine tasks are eliminated, new forms of expertise and oversight become essential. As COM2 stated: \enquote{My hypothesis and hope is that many things that don't create much added value will no longer require human resources and that the role of HR managers or all people who work in HR will focus more on value-adding things, such as building trust with employees, \textelp{} and also be able to turn business strategies into people strategies} (COM2). Ultimately, AI-augmented HR is intended to serve as a catalyst for a new understanding of value creation in the field. A strategic focus on value streams within organisational processes reflects an evolving vision of HR augmentation -- one that highlights the importance of talent for a company’s future success. This approach, however, remains ambiguous, positioned between economic imperatives of value creation and the interpersonal nature of HR work, which does not always align neatly with an economic logic.

\section{Discussion}\label{sec:discussion}
Our results demonstrate that the adoption of AI in HRM is not a simple, one-dimensional shift. Instead, it is a multifaceted process shaped by interconnected technological, organisational, and professional factors. First, our empirical findings show that AI tools -- from generative chatbots to predictive analytics systems -- are used across a range of HR functions. The scope of these applications depends largely on the availability of reliable data and the specific strategic objectives of each organisation. While some companies use AI to streamline routine administrative tasks and enhance decision-making, others are more cautious, often due to fragmented data infrastructures, limited digital readiness, or conflicting internal priorities.

Overall, our findings contribute to the theory of HRM digitalisation by (a) distinguishing between low-threshold generative applications and high-investment predictive analytics, (b) identifying co-determination as a distinctive institutional factor and catalyst in the German context, and (c) illustrating how AI is reshaping the professional identity of HR managers. As a result, HR professionals require new competencies in data governance, ethical auditing, and human-machine collaboration, while engaging in more focused interpersonal work. Ironically, AI-augmented HR could become even more people-centred in practice, but relies substantially on automation effects of AI.

First, we observe ambivalence with regard to RQ1 -- \textit{How do HR managers perceive and respond to the ongoing transformation driven by AI?} Practitioners appreciate AI’s capacity to relieve them of monotonous tasks, hoping this will free them to focus on strategic advisory, coaching, and relationship-building. However, many also admit having limited technical literacy and express concerns about opaque algorithmic decision-making (\enquote{black boxes}, \cite{Ajunwa.2020}), data biases, and potential legal liabilities under GDPR and the AI Act. As a result, HR professionals frequently emphasise maintaining a \enquote{human-in-the-loop} approach, utilising AI for preliminary screening or recommendations, but reserving final decisions for human judgment. This position can be viewed in two ways: as an ethical justification -- ensuring that machines do not make decisions about people -- and as a legal necessity, since regulatory frameworks would require it.

Second, for RQ2 -- \textit{What are the defining characteristics of organisational structures that employ AI in HRM} -- our data highlight the pivotal influence of organisational context. Larger, technology-driven, and internationally active firms with centralised HR information systems demonstrate greater readiness for AI, supported by executive buy-in, dedicated analytics teams, and standardised data \parencite[in line with][]{Arora.2024a, Chowdhury.2023}. Conversely, decentralised structures, fragmented systems, and skill gaps within HR departments hinder advanced analytics adoption. Furthermore, Germany’s co-determination framework adds an important layer of governance: works councils frequently demand transparency and hold veto rights over performance monitoring and assessment tools, ideally directing AI initiatives towards negotiated, participatory processes. Nevertheless, in practice, the concept of AI often remains intentionally ambiguous, which affects information flows in co-determination and participatory processes. Thus, AI narratives can serve multiple purposes -- selling products, justifying its strategic utility, managing high volumes, or raising critical issues for co-determination -- making its integration, contestation, or shaping an act of organisational storytelling \parencite{McCartney.2024}.

Additionally, some organisations pursue AI investments primarily to reduce costs and achieve rationalisation, while the much-promoted analytical, predictive, or prescriptive capabilities of AI related to workforce development or employee experience remain secondary. This pattern reflects earlier IoT and Industry 4.0 initiatives in Germany, which also focused mainly on efficiency gains \parencite{Kalff.2020, Butollo.2020}. This is of particular interest for our hypothesis about AI-augmented HR. Although the promoted analytical and predictive insights of HR analytics could support the case for AI-augmented HR work, in practice, the prevailing emphasis on rationalisation and efficiency does not genuinely “augment” HR work \parencite{Coron.2024}. Under this primarily efficiency-driven approach, HR may become increasingly transactional, relying on AI-based tools to accelerate tasks and facilitate short-term decisions \parencite{Malik.2023}.

Third, regarding RQ3 -- \textit{In what ways do AI tools augment HRM processes and work?} -- we find that companies integrate AI in two distinct ways. Generative models and conversational agents (such as LLM-based chatbots) are widely used to automate routine text generation, handle candidate inquiries, and synthesise documents \parencite{Raman.2024}. In this context, the hypothesis of AI-augmented HR work is well founded: the aim is to eliminate simple, repetitive administrative tasks and shift HR activities towards more meaningful, interpersonal, and value-oriented functions. In our research, AI augmentation supports core HR values such as people orientation, talent development, and human interaction. Several interviewees shared examples where efficiency gains from automation were redirected to enhance employee relations, well-being, and organisational culture. This creates a specific narrative: If AI alleviates administrative burdens, HR can devote greater attention to its relational and developmental roles. Nevertheless, the actual promises of augmentation through HR analytics, such as in-depth analytical insights, predictions or prescriptive action, play no role in the actual practice of enhanced HR work \parencite{Wirges.2023}. Our findings align with studies suggesting a shift in the professional identity of HRM towards greater strategic importance for organisations, as well as the associated complexities, such as a new reliance on data analytics and expertise distributed across various professional groups within organisations \parencite{Loscher.2022, Diefenhardt.2024}.

The limitations of our study include its cross-sectional design, its focus on Germany (where co-determination plays a pivotal role), and its reliance on self-reported survey data for quantitative measures. Further research is needed to examine how these technological strategies align with the ways companies assess the success of AI projects -- whether through short-term returns and reductions in headcount, or through broader outcomes such as improved employee retention and sustainable talent development. Moreover, comparative case studies of HR work with and without AI implementation could provide deeper insights into the informal, affective, and relational aspects of HR management under different technological conditions. Future research could also employ longitudinal or cross-jurisdictional approaches -- comparing settings with varying labour-law regimes -- to explore how institutional differences affect the adoption of AI in HRM.

\section{Conclusion}\label{sec:conclusion}
This paper has examined the landscape of AI-augmented HRM in Germany, revealing a broad spectrum of applications -- from readily accessible generative text tools and chatbots to more complex predictive analytics systems. Our study sheds light on the multifaceted realities of AI adoption in German HRM practices. While the integration of AI technologies can deliver substantial efficiencies and promote data-driven decision-making, these benefits are moderated by ongoing challenges related to data quality, organisational readiness, and ethical oversight. We contribute to the question whether AI will automate or augment human labour in HR and found that AI-augmented HR grounds in automation of undesired tasks that HR managers perceive as more resources for relevant, value creating, and people oriented tasks. AI augments HR work by enabling a greater focus on strategic responsibilities. At the same time, it demands that HR professionals expand their skill sets, developing not only traditional interpersonal abilities but also technical expertise and AI literacy.

Our study advances critical scholarship on AI in HRM by clarifying how organisational size, culture, co-determination rights, and strategic goals jointly shape the adoption of AI. It also traces the evolving skill requirements and identity work of HR professionals in an AI-enabled environment. By examining technological affordances, organisational conditions, and human agency together, future research can develop a nuanced understanding of AI-augmented HRM -- highlighting the dynamic interplay among technology, organisational context, and professional identity. As HR continues to evolve in the digital era, our findings underscore the importance of AI governance, ongoing competency development, and participatory oversight to ensure that AI serves as a catalyst for both operational efficiency and human-centred strategic value.

%\backmatter
\section*{Financial disclosure}
The project TranKI -- Standards for transparent AI was funded by the Hans Böckler Foundation, Düsseldorf, Germany. Grant No.: 2022-797-2, period: 10/2023 -- 09/2026.
\section*{Conflict of interest}
The authors declare no potential conflict of interests.

\printbibliography

@article{AfMalmborg.2023,
	title = {Discursive Framing and Organizational Venues: Mechanisms of Artificial Intelligence Policy Adoption},
	shorttitle = {Discursive Framing and Organizational Venues},
	author = {Af Malmborg, Frans and Trondal, Jarle},
	date = {2023-03},
	journaltitle = {International Review of Administrative Sciences},
	shortjournal = {Int. Rev. Adm. Sci.},
	volume = {89},
	number = {1},
	pages = {39--58},
	doi = {10.1177/00208523211007533},
	langid = {english}
}

@article{Ajunwa.2020,
	title = {The “Black Box” at Work},
	author = {Ajunwa, Ifeoma},
	date = {2020-01-01T00:00:00},
	journaltitle = {Big Data \& Society},
	shortjournal = {Big Data Soc.},
	volume = {7},
	number = {2},
	pages = {2053951720966181},
	doi = {10.1177/2053951720938093}
}

@article{Angrave.2016,
	title = {{{HR}} and Analytics: Why {{HR}} Is Set to Fail the Big Data Challenge},
	shorttitle = {{{HR}} and Analytics},
	author = {Angrave, David and Charlwood, Andy and Kirkpatrick, Ian and Lawrence, Mark and Stuart, Mark},
	date = {2016-01},
	journaltitle = {Human Resource Management Journal},
	shortjournal = {Hum. Resour. Manag. J.},
	volume = {26},
	number = {1},
	pages = {1--11},
	doi = {10.1111/1748-8583.12090},
	langid = {english}
}

@incollection{Arora.2021,
	title = {{{HR Analytics}} and {{Artificial Intelligence-Transforming Human Resource Management}}},
	booktitle = {2021 {{International Conference}} on {{Decision Aid Sciences}} and {{Application}} ({{DASA}}): 7-8 {{Dec}}. 2021},
	author = {Arora, Meenal and Prakash, Anshika and Mittal, Amit and Singh, Swati},
	date = {2021-01-01T00:00:00},
	pages = {288--293},
	publisher = {IEEE},
	location = {Piscataway},
	doi = {10.1109/DASA53625.2021.9682325},
	isbn = {978-1-6654-1634-4}
}

@article{Arora.2024a,
	title = {Adoption of {{HR}} Analytics for Future-Proof Decision Making: Role of Attitude toward Artificial Intelligence as a Moderator},
	shorttitle = {Adoption of {{HR}} Analytics for Future-Proof Decision Making},
	author = {Arora, Simple and Chaudhary, Priya and Singh, Reetesh K.},
	date = {2024-09-06},
	journaltitle = {International Journal of Organizational Analysis},
	shortjournal = {Int. J. Organ. Anal.},
	doi = {10.1108/IJOA-03-2024-4392},
	langid = {english}
}

@article{Ayling.2022,
	title = {Putting {{AI Ethics}} to {{Work}}: {{Are}} the {{Tools}} Fit for {{Purpose}}?},
	author = {Ayling, Jacqui and Chapman, Adriane},
	date = {2022-01-01T00:00:00},
	journaltitle = {AI and Ethics},
	shortjournal = {AI Ethics},
	volume = {2},
	number = {3},
	pages = {405--429},
	doi = {10.1007/s43681-021-00084-x}
}

@article{Bader.2019,
	title = {Algorithmic Decision-Making? {{The}} User Interface and Its Role for Human Involvement in Decisions Supported by Artificial Intelligence},
	author = {Bader, Verena and Kaiser, Stephan},
	date = {2019-01-01T00:00:00},
	journaltitle = {Organization},
	volume = {26},
	number = {5},
	pages = {655--672},
	doi = {10.1177/1350508419855714}
}

@article{Baldegger.2020,
	title = {Correlation between {{Entrepreneurial Orientation}} and {{Implementation}} of {{AI}} in {{Human Resources Management}}},
	author = {Baldegger, Rico and Caon, Maurizio and Sadiku, Kreshnik},
	date = {2020},
	journaltitle = {Technology Innovation Management Review},
	shortjournal = {Technol. Innov. Manag. Rev.},
	volume = {10},
	number = {4},
	pages = {72--79},
	url = {https://timreview.ca/article/1348}
}

@article{Bankins.2024,
	title = {A Multilevel Review of Artificial Intelligence in Organizations: {{Implications}} for Organizational Behavior Research and Practice},
	shorttitle = {A Multilevel Review of Artificial Intelligence in Organizations},
	author = {Bankins, Sarah and Ocampo, Anna Carmella and Marrone, Mauricio and Restubog, Simon Lloyd D. and Woo, Sang Eun},
	date = {2024-02},
	journaltitle = {Journal of Organizational Behavior},
	shortjournal = {J. Organ. Behav.},
	volume = {45},
	number = {2},
	pages = {159--182},
	doi = {10.1002/job.2735},
	langid = {english}
}

@article{Barrett.2013,
	title = {Envisioning {{E-HRM}} and Strategic {{HR}}: {{Taking}} Seriously Identity, Innovative Practice, and Service},
	shorttitle = {Envisioning {{E-HRM}} and Strategic {{HR}}},
	author = {Barrett, Michael and Oborn, Eivor},
	date = {2013-09},
	journaltitle = {The Journal of Strategic Information Systems},
	shortjournal = {J. Strateg. Inf. Syst.},
	volume = {22},
	number = {3},
	pages = {252--256},
	doi = {10.1016/j.jsis.2013.07.002},
	langid = {english}
}

@article{Bechter.2022,
	title = {The Role of the Capability, Opportunity, and Motivation of Firms for Using Human Resource Analytics to Monitor Employee Performance: {{A}} Multi‐level Analysis of the Organisational, Market, and Country Context},
	shorttitle = {The Role of the Capability, Opportunity, and Motivation of Firms for Using Human Resource Analytics to Monitor Employee Performance},
	author = {Bechter, Barbara and Brandl, Bernd and Lehr, Alex},
	date = {2022-11},
	journaltitle = {New Technology, Work and Employment},
	shortjournal = {New Technol. Work Employ.},
	volume = {37},
	number = {3},
	pages = {398--424},
	doi = {10.1111/ntwe.12239},
	langid = {english}
}

@article{Bohmer.2023,
	title = {Critical Exploration of {{AI-driven HRM}} to Build up Organizational Capabilities},
	author = {Böhmer, Nicole and Schinnenburg, Heike},
	date = {2023-07-24},
	journaltitle = {Employee Relations: The International Journal},
	shortjournal = {Empl. Relat. Int. J.},
	volume = {45},
	number = {5},
	pages = {1057--1082},
	doi = {10.1108/ER-04-2022-0202},
	langid = {english}
}

@incollection{Bohnsack.2010b,
	title = {Documentary {{Method}} and {{Group Discussions}}},
	booktitle = {Qualitative {{Analysis}} and {{Documentary Method}}: {{In International Educational Research}}},
	author = {Bohnsack, Ralf},
	editor = {Bohnsack, Ralf and Pfaff, Nicolle and Weller, Wivian},
	date = {2010-01-20},
	pages = {99--124},
	publisher = {Verlag Barbara Budrich},
	doi = {10.3224/86649236},
	isbn = {978-3-86649-236-3}
}

@article{Bolander.2013,
	title = {How {{Employee Selection Decisions}} Are {{Made}} in {{Practice}}},
	author = {Bolander, Pernilla and Sandberg, Jörgen},
	date = {2013-03},
	journaltitle = {Organization Studies},
	shortjournal = {Organ. Stud.},
	volume = {34},
	number = {3},
	pages = {285--311},
	doi = {10.1177/0170840612464757},
	langid = {english}
}

@article{Bondarouk.2016,
	title = {Conceptualising the Future of {{HRM}} and Technology Research},
	author = {Bondarouk, Tanya and Brewster, Chris},
	date = {2016-01-01T00:00:00},
	journaltitle = {The International Journal of Human Resource Management},
	shortjournal = {Int. J. Hum. Resour. Manag.},
	volume = {27},
	number = {21},
	pages = {2652--2671},
	doi = {10.1080/09585192.2016.1232296}
}

@book{Bryman.2011,
	title = {Business {{Research Methods}}},
	author = {Bryman, Alan and Bell, Emma},
	date = {2011-01-01T00:00:00},
	edition = {3},
	publisher = {Oxford University Press},
	location = {Oxford},
	isbn = {978-0-19-958340-9},
	pagetotal = {765}
}

@article{Budhwar.2022,
	title = {Artificial {{Intelligence}} – {{Challenges}} and {{Opportunities}} for International {{HRM}}: {{A Review}} and {{Research Agenda}}},
	author = {Budhwar, Pawan and Malik, Ashish and Silva, M. T. Thedushika and Thevisuthan, Praveena},
	date = {2022-01-01T00:00:00},
	journaltitle = {The International Journal of Human Resource Management},
	shortjournal = {Int. J. Hum. Resour. Manag.},
	volume = {33},
	number = {6},
	pages = {1065--1097},
	doi = {10.1080/09585192.2022.2035161}
}

@article{Budhwar.2023,
	title = {Human {{Resource Management}} in the {{Age}} of Generative {{Artificial Intelligence}}: {{Perspectives}} and Research Directions on {{ChatGPT}}},
	author = {Budhwar, Pawan and Chowdhury, Soumyadeb and Wood, Geoffrey and Aguinis, Herman and Bamber, Greg J. and Beltran, Jose R. and Boselie, Paul and Lee Cooke, Fang and Decker, Stephanie and DeNisi, Angelo and Dey, Prasanta Kumar and Guest, David and Knoblich, Andrew J. and Malik, Ashish and Paauwe, Jaap and Papagiannidis, Savvas and Patel, Charmi and Pereira, Vijay and Ren, Shuang and Rogelberg, Steven and Saunders, Mark N. K. and Tung, Rosalie L. and Varma, Arup},
	date = {2023-01-01T00:00:00},
	journaltitle = {Human Resource Management Journal},
	shortjournal = {Hum. Resour. Manag. J.},
	volume = {33},
	number = {3},
	pages = {606--659},
	doi = {10.1111/1748-8583.12524}
}

@incollection{Butollo.2020,
	title = {From {{Lean Production}} to {{Industrie}} 4.0: {{More Autonomy}} for {{Employees}}?},
	booktitle = {Digitalization in {{Industry}}: {{Work}} and {{Organisation}} -- between {{Domination}} and {{Emancipation}}},
	author = {Butollo, Florian and Jürgens, Ulrich and Krzywdzinski, Martin},
	editor = {Meyer, Uli and Schaupp, Simon and Seibt, David},
	date = {2019-01-01T00:00:00},
	pages = {61--80},
	publisher = {Palgrave Macmillan},
	location = {Basingstoke; New York},
	isbn = {978-3-030-28257-8}
}

@article{Cayrat.2023,
	title = {The Roles of the {{HR}} Function: {{A}} Systematic Review of Tensions, Continuity and Change},
	shorttitle = {The Roles of the {{HR}} Function},
	author = {Cayrat, Charles and Boxall, Peter},
	date = {2023-12},
	journaltitle = {Human Resource Management Review},
	shortjournal = {Hum. Resour. Manag. Rev.},
	volume = {33},
	number = {4},
	pages = {100984},
	doi = {10.1016/j.hrmr.2023.100984},
	langid = {english}
}

@article{Charlwood.2022,
	title = {Can {{HR}} Adapt to the Paradoxes of Artificial Intelligence?},
	author = {Charlwood, Andy and Guenole, Nigel},
	date = {2022-01-01T00:00:00},
	journaltitle = {Human Resource Management Journal},
	shortjournal = {Hum. Resour. Manag. J.},
	volume = {32},
	number = {4},
	pages = {729--742},
	doi = {10.1111/1748-8583.12433},
	pagetotal = {14}
}

@article{Chowdhury.2023,
	title = {Unlocking the Value of Artificial Intelligence in Human Resource Management through {{AI}} Capability Framework},
	author = {Chowdhury, Soumyadeb and Dey, Prasanta and Joel-Edgar, Sian and Bhattacharya, Sudeshna and Rodriguez-Espindola, Oscar and Abadie, Amelie and Truong, Linh},
	date = {2023-01-01T00:00:00},
	journaltitle = {Human Resource Management Review},
	shortjournal = {Hum. Resour. Manag. Rev.},
	volume = {33},
	number = {1},
	pages = {100899},
	doi = {10.1016/j.hrmr.2022.100899}
}

@book{Ciesla.2024,
	title = {The {{Book}} of {{Chatbots}}: {{From ELIZA}} to {{ChatGPT}}},
	shorttitle = {The {{Book}} of {{Chatbots}}},
	author = {Ciesla, Robert},
	date = {2024},
	publisher = {Springer Nature Switzerland},
	location = {Cham},
	doi = {10.1007/978-3-031-51004-5},
	isbn = {978-3-031-51003-8},
	langid = {english}
}

@article{Coolen.2023,
	title = {Understanding the Adoption and Institutionalization of Workforce Analytics: {{A}} Systematic Literature Review and Research Agenda},
	shorttitle = {Understanding the Adoption and Institutionalization of Workforce Analytics},
	author = {Coolen, Patrick and Van Den Heuvel, Sjoerd and Van De Voorde, Karina and Paauwe, Jaap},
	date = {2023-12},
	journaltitle = {Human Resource Management Review},
	shortjournal = {Hum. Resour. Manag. Rev.},
	volume = {33},
	number = {4},
	pages = {100985},
	doi = {10.1016/j.hrmr.2023.100985},
	langid = {english}
}

@article{Coron.2022,
	title = {Quantifying Human Resource Management: A Literature Review},
	author = {Coron, Clotilde},
	date = {2022-01-01T00:00:00},
	journaltitle = {Personnel Review},
	shortjournal = {Pers. Rev.},
	volume = {51},
	number = {4},
	pages = {1386--1409},
	doi = {10.1108/PR-05-2020-0322}
}

@article{Coron.2024,
	title = {How to Do {{HRM}} with {{Numbers}}? {{A}} Performative {{Lens}} on {{HR Metrics}}, {{HR Analytics}} and {{HR Algorithms}}},
	author = {Coron, Clotilde and Scheibmayr, Isabella and Lescoat, Pierre},
	date = {2024-01-01T00:00:00},
	journaltitle = {New Technology, Work and Employment},
	shortjournal = {New Technol. Work Employ.},
	doi = {10.1111/ntwe.12306},
	volumes = {ntwe.12306}
}

@article{Das.2026,
	title = {From {{Automation}} to {{Augmentation}}: {{A Bibliometric}} and {{Thematic Review}} of {{Artificial Intelligence}} in {{Human Resource Management}}},
	shorttitle = {From {{Automation}} to {{Augmentation}}},
	author = {Das, Tilak Ch and Neog, Aparajita and Amin, Md Ruhul and Ahmed, Fojail},
	date = {2026-05-08},
	journaltitle = {International Review of Management and Marketing},
	shortjournal = {Int. Rev. Manag. Mark.},
	volume = {16},
	number = {4},
	pages = {17--32},
	doi = {10.32479/irmm.23035}
}

@article{Deng.2024,
	title = {The {{Power}} of {{Precision}}: {{How Algorithmic Monitoring}} and {{Performance Management Enhances Employee Workplace Well}}‐{{Being}}},
	shorttitle = {The {{Power}} of {{Precision}}},
	author = {Deng, Hui and Lu, Ying and Fan, Di and Liu, Wei and Xia, Yuhuan},
	date = {2024-12-25},
	journaltitle = {New Technology, Work and Employment},
	shortjournal = {New Technol. Work Employ.},
	pages = {ntwe.12328},
	doi = {10.1111/ntwe.12328},
	langid = {english}
}

@article{Diefenhardt.2024,
	title = {‘{{In God We Trust}}. {{All Others Must Bring Data}}’: {{Unpacking}} the {{Influence}} of {{Human Resource Analytics}} on the {{Strategic Recognition}} of {{Human Resource Management}}},
	shorttitle = {‘{{In God We Trust}}. {{All Others Must Bring Data}}’},
	author = {Diefenhardt, Felix and Rapp, Marco L. and Bader, Verena and Mayrhofer, Wolfgang},
	date = {2024-11-27},
	journaltitle = {Human Resource Management Journal},
	shortjournal = {Hum. Resour. Manag. J.},
	pages = {1748-8583.12583},
	doi = {10.1111/1748-8583.12583},
	langid = {english}
}

@article{Drage.2022,
	title = {Does {{AI Debias Recruitment}}? {{Race}}, {{Gender}}, and {{AI}}'s \enquote{{{Eradication}} of {{Difference}}}},
	author = {Drage, Eleanor and Mackereth, Kerry},
	date = {2022-01-01T00:00:00},
	journaltitle = {Philosophy \& Technology},
	shortjournal = {Philos. Technol.},
	volume = {35},
	number = {4},
	eprint = {36246553},
	eprinttype = {pubmed},
	pages = {1--25},
	doi = {10.1007/s13347-022-00543-1},
	volumes = {89}
}

@book{Drisko.2016,
	title = {Content Analysis},
	author = {Drisko, James W. and Maschi, Tina},
	date = {2016-01-01T00:00:00},
	publisher = {Oxford University Press},
	location = {Oxford; New York},
	pagetotal = {191}
}

@article{Du.2024,
	title = {Exploring {{Gender Bias}} and {{Algorithm Transparency}}: {{Ethical Considerations}} of {{AI}} in {{HRM}}},
	author = {Du, Jiaxing},
	date = {2024-01-01T00:00:00},
	journaltitle = {Journal of Theory and Practice of Management Science},
	shortjournal = {J. Theory Pract. Manag. Sci.},
	volume = {4},
	number = {03},
	pages = {36--43},
	doi = {10.53469/jtpms.2024.04(03).06}
}

@article{Dutta.2024,
	title = {The Machine/Human Agentic Impact on Practices in Learning and Development: A Study across {{MSME}}, {{NGO}} and {{MNC}} Organizations},
	shorttitle = {The Machine/Human Agentic Impact on Practices in Learning and Development},
	author = {Dutta, Debolina and Kannan Poyil, Anasha},
	date = {2024-05-13},
	journaltitle = {Personnel Review},
	shortjournal = {Pers. Rev.},
	volume = {53},
	number = {3},
	pages = {791--815},
	doi = {10.1108/PR-09-2022-0658},
	langid = {english}
}

@report{EuropeanCommission.2021,
	title = {Regulation of the {{European Parliament}} and of the {{Council}} Laying down Harmonised {{Rules}} on {{Artificial Intelligence}} ({{Artificial Intelligence Act}}) and Amending Certain {{Union}} Legislative {{Acts}}},
	author = {{European Commission}},
	date = {2021-01-01T00:00:00},
	location = {Brussels},
	url = {https://eur-lex.europa.eu/resource.html?uri=cellar:e0649735-a372-11eb-9585-01aa75ed71a1.0001.02/DOC_1&format=PDF},
	issue = {2021/0106 (COD)}
}

@report{EuropeanParliament.2016,
	title = {General {{Data Protection Regulation}}: {{GDPR}}},
	author = {{European Parliament}},
	date = {2016-01-01T00:00:00},
	journaltitle = {Official Journal of the European Union},
	number = {L119/1},
	url = {https://gdpr-info.eu/}
}

@article{Fabris.2024,
	title = {Fairness and {{Bias}} in {{Algorithmic Hiring}}: {{A Multidisciplinary Survey}}},
	author = {Fabris, Alessandro and Baranowska, Nina and Dennis, Matthew J. and Graus, David and Hacker, Philipp and Saldivar, Jorge and Zuiderveen Borgesius, Frederik and Biega, Asia J.},
	date = {2024-01-01T00:00:00},
	journaltitle = {ACM Transactions on Intelligent Systems and Technology},
	shortjournal = {ACM Trans. Intell. Syst. Technol.},
	doi = {10.1145/3696457},
	volumes = {3696457}
}

@article{Fenwick.2024,
	title = {The Critical Role of {{HRM}} in {{AI-driven}} Digital Transformation: A Paradigm Shift to Enable Firms to Move from {{AI}} Implementation to Human-Centric Adoption},
	shorttitle = {The Critical Role of {{HRM}} in {{AI-driven}} Digital Transformation},
	author = {Fenwick, Ali and Molnar, Gabor and Frangos, Piper},
	date = {2024-05-13},
	journaltitle = {Discover Artificial Intelligence},
	shortjournal = {Discov. Artif. Intell.},
	volume = {4},
	number = {1},
	pages = {34},
	doi = {10.1007/s44163-024-00125-4},
	langid = {english}
}

@article{Fernandez.2021,
	title = {Tackling the {{HR}} Digitalization Challenge: Key Factors and Barriers to {{HR}} Analytics Adoption},
	shorttitle = {Tackling the {{HR}} Digitalization Challenge},
	author = {Fernandez, Vicenc and Gallardo-Gallardo, Eva},
	date = {2021-01-17},
	journaltitle = {Competitiveness Review: An International Business Journal},
	shortjournal = {Compet. Rev. Int. Bus. J.},
	volume = {31},
	number = {1},
	pages = {162--187},
	doi = {10.1108/CR-12-2019-0163},
	langid = {english}
}

@article{Fourcade.2016,
	title = {Seeing like a Market},
	author = {Fourcade, Marion and Healy, Kieran},
	date = {2016-12-23},
	journaltitle = {Socio-Economic Review},
	shortjournal = {Socio-Econ. Rev.},
	volume = {15},
	number = {1},
	pages = {9--29},
	doi = {10.1093/ser/mww033},
	langid = {english}
}

@incollection{Glaser.2009,
	title = {On {{Interviewing}} “{{Good}}” and “{{Bad}}” {{Experts}}},
	booktitle = {Interviewing Experts: {{Methodology}} and {{Practice}}},
	author = {Gläser, Jochen and Laudel, Grit},
	editor = {Bogner, Alexander and Littig, Beate and Menz, Wolfgang},
	date = {2009-01-01T00:00:00},
	pages = {117--137},
	publisher = {Palgrave Macmillan},
	location = {Basingstoke; New York},
	doi = {10.1057/9780230244276_6},
	isbn = {978-0-230-22019-5}
}

@article{Jatoba.2023,
	title = {Intelligent Human Resources for the Adoption of Artificial Intelligence: A Systematic Literature Review},
	shorttitle = {Intelligent Human Resources for the Adoption of Artificial Intelligence},
	author = {Jatobá, Mariana Namen and Ferreira, João J. and Fernandes, Paula Odete and Teixeira, João Paulo},
	date = {2023-12-11},
	journaltitle = {Journal of Organizational Change Management},
	shortjournal = {J. Organ. Change Manag.},
	volume = {36},
	number = {7},
	pages = {1099--1124},
	doi = {10.1108/JOCM-03-2022-0075},
	langid = {english}
}

@incollection{Kalff.2020,
	title = {Labor {{Democracy}} in {{Digitalizing Industries}}: {{Emancipating}} or »sandboxing« {{Participation}} in {{Discourses}} on {{Technology}} and New {{Forms}} of {{Work}}?},
	booktitle = {Digitalization in {{Industry}}: {{Work}} and {{Organisation}} -- between {{Domination}} and {{Emancipation}}},
	author = {Kalff, Yannick},
	editor = {Meyer, Uli and Schaupp, Simon and Seibt, David},
	date = {2019-01-01T00:00:00},
	pages = {29--60},
	publisher = {Palgrave Macmillan},
	location = {Basingstoke; New York},
	doi = {10.1007/978-3-030-28258-5_2},
	isbn = {978-3-030-28257-8}
}

@article{Kalff.2022b,
	title = {Beschäftigtenvorbehalte Gegen Digitale {{Assistenzsysteme}}: {{Konfliktquellen}} Und Partizipative {{Technikgestaltung}} Im Soziotechnischen {{System}}},
	author = {Kalff, Yannick and Kutlu, Yalçın},
	date = {2022-01-01T00:00:00},
	journaltitle = {Arbeit},
	volume = {31},
	number = {4},
	pages = {377--398},
	doi = {10.1515/arbeit-2022-0022},
	pagetotal = {22}
}

@article{Kalff.2023,
	title = {Die {{Real-Utopie}} Soziokratischer Und Demokratischer {{Wirtschaftsorganisationen}}: {{Ergebnisse}} Einer Qualitativ-Explorativen {{Untersuchung}}},
	author = {Kalff, Yannick},
	date = {2023-01-01T00:00:00},
	journaltitle = {Z’GuG -- Zeitschrift für Gemeinwirtschaft und Gemeinwohl},
	shortjournal = {Z’GuG -- Z. Für Gemeinwirtsch. Gemeinwohl},
	volume = {46},
	number = {3},
	pages = {284--301},
	doi = {10.5771/2701-4193-2023-3-284},
	pagetotal = {18}
}

@inproceedings{Kalff.2025c,
	title = {Explained, yet Misunderstood: {{How AI Literacy}} Shapes {{HR Managers}}' Interpretation of {{User Interfaces}} in {{Recruiting Recommender Systems}}},
	shorttitle = {Explained, yet Misunderstood},
	booktitle = {Proceedings of the 5th {{Workshop}} on {{Recommender Systems}} for {{Human Resources}} ({{RecSys}} in {{HR}} 2025)},
	author = {Kalff, Yannick and Simbeck, Katharina},
	editor = {Kaya, Mesut and Bogers, Toine and Bied, Guillaume and Johnson, Chris and Decorte, Jens-Joris},
	date = {2025},
	volume = {4046},
	publisher = {CEUR},
	location = {Prague},
	url = {https://ceur-ws.org/Vol-4046/RecSysHR2025-paper_3.pdf},
	urldate = {2025-09-28},
	eventtitle = {{{RecSys}} in {{HR}} 2025},
	version = {1}
}

@article{Kelan.2024,
	title = {Algorithmic Inclusion: {{Shaping}} the Predictive Algorithms of Artificial Intelligence in Hiring},
	author = {Kelan, Elisabeth K.},
	date = {2024-01-01T00:00:00},
	journaltitle = {Human Resource Management Journal},
	shortjournal = {Hum. Resour. Manag. J.},
	volume = {34},
	number = {3},
	pages = {694--707},
	doi = {10.1111/1748-8583.12511}
}

@article{Kim.2021,
	title = {Sixty Years of Research on Technology and Human Resource Management: {{Looking}} Back and Looking Forward},
	author = {Kim, Sunghoon and Wang, Ying and Boon, Corine},
	date = {2021-01-01T00:00:00},
	journaltitle = {Human Resource Management},
	shortjournal = {Hum. Resour. Manage.},
	volume = {60},
	number = {1},
	pages = {229--247},
	doi = {10.1002/hrm.22049}
}

@article{Kochling.2021,
	title = {Highly {{Accurate}}, {{But Still Discriminatory}}: {{A Fairness Evaluation}} of {{Algorithmic Video Analysis}} in the {{Recruitment Context}}},
	author = {Köchling, Alina and Riazy, Shirin and Wehner, Marius Claus and Simbeck, Katharina},
	date = {2021-01-01T00:00:00},
	journaltitle = {Business \& Information Systems Engineering},
	shortjournal = {Bus. Inf. Syst. Eng.},
	volume = {63},
	number = {1},
	pages = {39--54},
	doi = {10.1007/s12599-020-00673-w}
}

@article{Kochling.2024,
	title = {This ({{AI}})n’t Fair? {{Employee}} Reactions to Artificial Intelligence ({{AI}}) in Career Development Systems},
	author = {Köchling, Alina and Wehner, Marius Claus and Ruhle, Sascha Alexander},
	date = {2024-01-01T00:00:00},
	journaltitle = {Review of Managerial Science},
	shortjournal = {Rev. Manag. Sci.},
	doi = {10.1007/s11846-024-00789-3},
	pagetotal = {35}
}

@article{Korherr.2023,
	title = {The Role of Management in Fostering Analytics: The Shift from Intuition to Analytics-Based Decision-Making},
	shorttitle = {The Role of Management in Fostering Analytics},
	author = {Korherr, Philipp and Kanbach, Dominik K. and Kraus, Sascha and Jones, Paul},
	date = {2023-07-26},
	journaltitle = {Journal of Decision Systems},
	shortjournal = {J. Decis. Syst.},
	volume = {32},
	number = {3},
	pages = {600--616},
	doi = {10.1080/12460125.2022.2062848},
	langid = {english}
}

@book{Kuckartz.2023,
	title = {Qualitative {{Content Analysis}}: {{Methods}}, {{Practice}} and {{Software}}},
	author = {Kuckartz, Udo and Rädiker, Stefan},
	date = {2023-01-01T00:00:00},
	edition = {2},
	publisher = {SAGE},
	location = {London; Thousand Oaks; New Delhi; Singapore},
	isbn = {978-1-5296-0913-4},
	pagetotal = {236}
}

@article{Lehmann.2024,
	title = {{{HR}} Analytics: {{A}} Centralizing or Decentralizing Force?},
	shorttitle = {{{HR}} Analytics},
	author = {Lehmann, Johannes},
	date = {2024},
	publisher = {WWZ},
	doi = {10.5451/UNIBAS-EP96815}
}

@incollection{Lippert.2024,
	title = {Artificial {{Intelligence}} and the {{Future}} of Managerial {{Roles}}: {{A}} Theoretical {{Review}}},
	booktitle = {European {{Conference}} on {{Information Systems}} ({{ECIS}}) 2024},
	author = {Lippert, Isabell},
	namea = {ECIS},
	nameatype = {collaborator},
	date = {2024-01-01T00:00:00}
}

@inproceedings{Lodra.2024,
	title = {The {{Impact}} of {{Artificial Intelligence}} on {{Recruitment}} and {{Selection}} for {{Human Resource Management}}: {{A Systematic Literature Review}}},
	shorttitle = {The {{Impact}} of {{Artificial Intelligence}} on {{Recruitment}} and {{Selection}} for {{Human Resource Management}}},
	booktitle = {2024 {{International Conference}} on {{ICT}} for {{Smart Society}} ({{ICISS}})},
	author = {Lodra, Raynald Samuel and Padhana, Tyaga and Kristin, Desi Maya},
	date = {2024-09-04},
	pages = {1--6},
	publisher = {IEEE},
	location = {Bandung, Indonesia},
	doi = {10.1109/ICISS62896.2024.10751529},
	eventtitle = {2024 {{International Conference}} on {{ICT}} for {{Smart Society}} ({{ICISS}})}
}

@incollection{Loscher.2022,
	title = {Augmenting a {{Profession}}: {{How Data Analytics}} Is {{Transforming Human Resource Management}}},
	shorttitle = {Augmenting a {{Profession}}},
	booktitle = {Research in the {{Sociology}} of {{Organizations}}},
	author = {Loscher, Georg and Bader, Verena},
	editor = {Gegenhuber, Thomas and Logue, Danielle and Hinings, C.R. (Bob) and Barrett, Michael},
	date = {2022-09-23},
	pages = {87--110},
	publisher = {Emerald Publishing Limited},
	doi = {10.1108/S0733-558X20220000083004},
	isbn = {978-1-80262-222-5}
}

@article{Loscher.2023,
	title = {Creating Accountability through {{HR}} Analytics – {{An}} Audit Society Perspective},
	author = {Loscher, Georg and Bader, Verena},
	date = {2023-12},
	journaltitle = {Human Resource Management Review},
	shortjournal = {Hum. Resour. Manag. Rev.},
	volume = {33},
	number = {4},
	pages = {100974},
	doi = {10.1016/j.hrmr.2023.100974},
	langid = {english}
}

@article{Malik.2023,
	title = {Artificial Intelligence ({{AI}})-Assisted {{HRM}}: {{Towards}} an Extended Strategic Framework},
	author = {Malik, Ashish and Budhwar, Pawan and Kazmi, Bahar Ali},
	date = {2023-01-01T00:00:00},
	journaltitle = {Human Resource Management Review},
	shortjournal = {Hum. Resour. Manag. Rev.},
	volume = {33},
	number = {1},
	pages = {100940},
	doi = {10.1016/j.hrmr.2022.100940}
}

@article{Marler.2017,
	title = {An Evidence-Based Review of {{HR Analytics}}},
	author = {Marler, Janet H. and Boudreau, John W.},
	date = {2017-01-02},
	journaltitle = {The International Journal of Human Resource Management},
	shortjournal = {Int. J. Hum. Resour. Manag.},
	volume = {28},
	number = {1},
	pages = {3--26},
	doi = {10.1080/09585192.2016.1244699},
	langid = {english}
}

@article{McCartney.2022,
	title = {Promise versus Reality: A Systematic Review of the Ongoing Debates in People Analytics},
	shorttitle = {Promise versus Reality},
	author = {McCartney, Steven and Fu, Na},
	date = {2022-04-20},
	journaltitle = {Journal of Organizational Effectiveness: People and Performance},
	shortjournal = {J. Organ. Eff. People Perform.},
	volume = {9},
	number = {2},
	pages = {281--311},
	doi = {10.1108/JOEPP-01-2021-0013},
	langid = {english}
}

@article{McCartney.2024,
	title = {Enacting People Analytics: {{Exploring}} the Direct and Complementary Effects of Analytical and Storytelling Skills},
	shorttitle = {Enacting People Analytics},
	author = {McCartney, Steven and Fu, Na},
	date = {2024-03},
	journaltitle = {Human Resource Management},
	shortjournal = {Hum. Resour. Manage.},
	volume = {63},
	number = {2},
	pages = {187--205},
	doi = {10.1002/hrm.22194},
	langid = {english}
}

@inproceedings{Medaglia.2022,
	title = {The Adoption of {{Artificial Intelligence}} in the Public Sector in {{Europe}}: Drivers, Features, and Impacts},
	shorttitle = {The Adoption of {{Artificial Intelligence}} in the Public Sector in {{Europe}}},
	booktitle = {Proceedings of the 15th {{International Conference}} on {{Theory}} and {{Practice}} of {{Electronic Governance}}},
	author = {Medaglia, Rony and Tangi, Luca},
	date = {2022-10-04},
	pages = {10--18},
	publisher = {ACM},
	location = {Guimarães Portugal},
	doi = {10.1145/3560107.3560110},
	eventtitle = {{{ICEGOV}} 2022: 15th {{International Conference}} on {{Theory}} and {{Practice}} of {{Electronic Governance}}},
	isbn = {978-1-4503-9635-6},
	langid = {english}
}

@incollection{Mer.2023,
	title = {Artificial {{Intelligence}} in {{Human Resource Management}}: {{Recent Trends}} and {{Research Agenda}}},
	booktitle = {Digital Transformation, Strategic Resilience, Cyber Security and Risk Management},
	author = {Mer, Akansha},
	editor = {Grima, Simon and Thalassinos, Eleftherios and Noja, Gratiela Georgiana and Stamataopoulos, Theodore V. and Vasiljeva, Tatjana and Volkova, Tatjana},
	date = {2023-01-01T00:00:00},
	series = {Contemporary {{Studies}} in {{Economic}} and {{Financial Analysis}}},
	edition = {First edition},
	volume = {Volume 111B},
	pages = {31--56},
	publisher = {Emerald Publishing Limited},
	location = {Bingley, U.K.},
	doi = {10.1108/S1569-37592023000111B003},
	isbn = {978-1-80455-262-9}
}

@article{Minbaeva.2021,
	title = {Disrupted {{HR}}?},
	author = {Minbaeva, Dana},
	date = {2021-12},
	journaltitle = {Human Resource Management Review},
	shortjournal = {Hum. Resour. Manag. Rev.},
	volume = {31},
	number = {4},
	pages = {100820},
	doi = {10.1016/j.hrmr.2020.100820},
	langid = {english}
}

@article{Muridzi.2024,
	title = {Artificial {{Intelligence}} in Transforming {{HRM Processes}} within {{Organizations}}},
	author = {Muridzi, Gibson and Dhliwayo, Shepherd and Isabelle, Diane A.},
	date = {2024-12},
	journaltitle = {International Journal of Innovation Management},
	shortjournal = {Int. J. Innov. Manag.},
	volume = {28},
	pages = {2440006},
	doi = {10.1142/S1363919624400061},
	issue = {09n10},
	langid = {english}
}

@book{Narayanan.2024,
	title = {{{AI Snake Oil}}: {{What Artificial Intelligence}} Can Do, What It Can't, and How to Tell the {{Difference}}},
	author = {Narayanan, Arvind and Kapoor, Sayash},
	date = {2024-01-01T00:00:00},
	publisher = {Princeton University Press},
	location = {Princeton},
	isbn = {978-0-691-24913-1}
}

@article{Nawaz.2024,
	title = {The Adoption of Artificial Intelligence in Human Resources Management Practices},
	author = {Nawaz, Nishad and Arunachalam, Hemalatha and Pathi, Barani Kumari and Gajenderan, Vijayakumar},
	date = {2024-04},
	journaltitle = {International Journal of Information Management Data Insights},
	shortjournal = {Int. J. Inf. Manag. Data Insights},
	volume = {4},
	number = {1},
	pages = {100208},
	doi = {10.1016/j.jjimei.2023.100208},
	langid = {english}
}

@article{Neumann.2024,
	title = {Exploring Artificial Intelligence Adoption in Public Organizations: A Comparative Case Study},
	shorttitle = {Exploring Artificial Intelligence Adoption in Public Organizations},
	author = {Neumann, Oliver and Guirguis, Katharina and Steiner, Reto},
	date = {2024-01-02},
	journaltitle = {Public Management Review},
	shortjournal = {Public Manag. Rev.},
	volume = {26},
	number = {1},
	pages = {114--141},
	doi = {10.1080/14719037.2022.2048685},
	langid = {english}
}

@article{Ogbeibu.2024,
	title = {Demystifying the {{Roles}} of Organisational Smart {{Technology}}, {{Artificial Intelligence}}, {{Robotics}} and {{Algorithms Capability}}: {{A Strategy}} for Green {{Human Resource Management}} and Environmental {{Sustainability}}},
	author = {Ogbeibu, Samuel and Emelifeonwu, Jude and Pereira, Vijay and Oseghale, Raphael and Gaskin, James and Sivarajah, Uthayasankar and Gunasekaran, Angappa},
	date = {2024-01-01T00:00:00},
	journaltitle = {Business Strategy and the Environment},
	shortjournal = {Bus. Strategy Environ.},
	volume = {33},
	number = {2},
	pages = {369--388},
	doi = {10.1002/bse.3495}
}

@article{Pan.2022,
	title = {The Adoption of Artificial Intelligence in Employee Recruitment: {{The}} Influence of Contextual Factors},
	shorttitle = {The Adoption of Artificial Intelligence in Employee Recruitment},
	author = {Pan, Yuan and Froese, Fabian and Liu, Ni and Hu, Yunyang and Ye, Maolin},
	date = {2022-03-26},
	journaltitle = {The International Journal of Human Resource Management},
	shortjournal = {Int. J. Hum. Resour. Manag.},
	volume = {33},
	number = {6},
	pages = {1125--1147},
	doi = {10.1080/09585192.2021.1879206},
	langid = {english}
}

@article{Panayotopoulou.2010,
	title = {Adoption of Electronic Systems in {{HRM}}: Is National Background of the Firm Relevant? {{Adoption}} of Electronic Systems in {{HRM}}},
	shorttitle = {Adoption of Electronic Systems in {{HRM}}},
	author = {Panayotopoulou, Leda and Galanaki, Eleanna and Papalexandris, Nancy},
	date = {2010-11},
	journaltitle = {New Technology, Work and Employment},
	shortjournal = {New Technol. Work Employ.},
	volume = {25},
	number = {3},
	pages = {253--269},
	doi = {10.1111/j.1468-005X.2010.00252.x},
	langid = {english}
}

@article{Pillai.2020,
	title = {Adoption of Artificial Intelligence ({{AI}}) for Talent Acquisition in {{IT}}/{{ITeS}} Organizations},
	author = {Pillai, Rajasshrie and Sivathanu, Brijesh},
	date = {2020-08-14},
	journaltitle = {Benchmarking: An International Journal},
	shortjournal = {Benchmarking Int. J.},
	volume = {27},
	number = {9},
	pages = {2599--2629},
	doi = {10.1108/BIJ-04-2020-0186},
	langid = {english}
}

@inproceedings{Poba-Nzaou.2022,
	title = {Understanding {{Artificial Intelligence Adoption Predictors}}: {{Empirical Insights}} from {{A Large-Scale Survey}}},
	shorttitle = {Understanding {{Artificial Intelligence Adoption Predictors}}},
	booktitle = {2022 {{International Conference}} on {{Information Management}} and {{Technology}} ({{ICIMTech}})},
	author = {Poba-Nzaou, Placide and Tchibozo, Anicet Sylvere},
	date = {2022-08-11},
	pages = {323--326},
	publisher = {IEEE},
	location = {Semarang, Indonesia},
	doi = {10.1109/ICIMTech55957.2022.9915214},
	eventtitle = {2022 {{International Conference}} on {{Information Management}} and {{Technology}} ({{ICIMTech}})},
	isbn = {978-1-6654-5090-4}
}

@article{Prikshat.2023,
	title = {{{AI-Augmented HRM}}: {{Literature}} Review and a Proposed Multilevel Framework for Future Research},
	shorttitle = {{{AI-Augmented HRM}}},
	author = {Prikshat, Verma and Islam, Mohammad and Patel, Parth and Malik, Ashish and Budhwar, Pawan and Gupta, Suraksha},
	date = {2023-08},
	journaltitle = {Technological Forecasting and Social Change},
	shortjournal = {Technol. Forecast. Soc. Change},
	volume = {193},
	pages = {122645},
	doi = {10.1016/j.techfore.2023.122645},
	langid = {english}
}

@article{Prikshat.2023a,
	title = {{{AI-augmented HRM}}: {{Antecedents}}, Assimilation and Multilevel Consequences},
	shorttitle = {{{AI-augmented HRM}}},
	author = {Prikshat, Verma and Malik, Ashish and Budhwar, Pawan},
	date = {2023-03},
	journaltitle = {Human Resource Management Review},
	shortjournal = {Hum. Resour. Manag. Rev.},
	volume = {33},
	number = {1},
	pages = {100860},
	doi = {10.1016/j.hrmr.2021.100860},
	langid = {english}
}

@article{Qamar.2021,
	title = {When {{Technology}} Meets {{People}}: {{The Interplay}} of {{Artificial Intelligence}} and {{Human Resource Management}}},
	author = {Qamar, Yusra and Agrawal, Rakesh Kumar and Samad, Taab Ahmad and Chiappetta Jabbour, Charbel Jose},
	date = {2021-01-01T00:00:00},
	journaltitle = {Journal of Enterprise Information Management},
	shortjournal = {J. Enterp. Inf. Manag.},
	volume = {34},
	number = {5},
	pages = {1339--1370},
	doi = {10.1108/JEIM-11-2020-0436}
}

@book{Radiker.2020,
	title = {Focused {{Analysis}} of {{Qualitative Interviews}} with {{MAXQDA}}},
	author = {Rädiker, Stefan and Kuckartz, Udo},
	date = {2020-01-01T00:00:00},
	publisher = {MAXQDA Press},
	location = {Berlin},
	doi = {10.36192/978-3-948768072},
	isbn = {978-3-948768-07-2}
}

@article{Raisch.2021,
	title = {Artificial {{Intelligence}} and {{Management}}: {{The Automation-Augmentation Paradox}}},
	author = {Raisch, Sebastian and Krakowski, Sebastian},
	date = {2021-01-01T00:00:00},
	journaltitle = {Academy of Management Review},
	shortjournal = {Acad. Manage. Rev.},
	volume = {46},
	number = {1},
	pages = {192--210},
	doi = {10.5465/amr.2018.0072}
}

@article{Raman.2024,
	title = {Evaluating Human Resources Management Literacy: {{A}} Performance Analysis of {{ChatGPT}} and Bard},
	author = {Raman, Raghu and Venugopalan, Murale and Kamal, Anju},
	date = {2024-01-01T00:00:00},
	journaltitle = {Heliyon},
	volume = {10},
	number = {5},
	eprint = {38486738},
	eprinttype = {pubmed},
	pages = {27026},
	doi = {10.1016/j.heliyon.2024.e27026}
}

@article{Revillod.2024,
	title = {Implementation of {{AI Recruitment Systems}} in {{Swiss HRM}}: {{The Importance}} of {{Technological}} and {{Organizational Factors}}},
	shorttitle = {Implementation of {{AI Recruitment Systems}} in {{Swiss HRM}}},
	author = {Revillod, Guillaume},
	date = {2024-11-24},
	journaltitle = {Journal of Human Resource Management - HR Advances and Developments},
	shortjournal = {J. Hum. Resour. Manag. - HR Adv. Dev.},
	volume = {2024},
	number = {2},
	pages = {95--122},
	doi = {10.46287/YDNH4362}
}

@inproceedings{Rukadikar.2023,
	title = {Adoption {{Of Artificial Intelligence In Talent Acquisition}}: {{The Need For The E-Business Environment}}},
	shorttitle = {Adoption {{Of Artificial Intelligence In Talent Acquisition}}},
	booktitle = {2023 8th {{International Conference}} on {{Business}} and {{Industrial Research}} ({{ICBIR}})},
	author = {Rukadikar, Aaradhana and Pandita, Deepika and Choudhary, Himani},
	date = {2023-05-18},
	pages = {228--232},
	publisher = {IEEE},
	location = {Bangkok, Thailand},
	doi = {10.1109/ICBIR57571.2023.10147592},
	eventtitle = {2023 8th {{International Conference}} on {{Business}} and {{Industrial Research}} ({{ICBIR}})},
	isbn = {979-8-3503-9964-6}
}

@book{Schellmann.2024,
	title = {The Algorithm: {{How AI}} Decides Who Gets Hired, Monitored, Promoted, and Fired and Why We Need to Fight Back Now},
	author = {Schellmann, Hilke},
	date = {2024-01-01T00:00:00},
	publisher = {Hachette Books},
	location = {New York},
	isbn = {978-0-306-82734-1}
}

@article{Simbeck.2019,
	title = {{{HR}} Analytics and {{Ethics}}},
	author = {Simbeck, Katharina},
	date = {2019-01-01T00:00:00},
	journaltitle = {IBM Journal of Research and Development},
	shortjournal = {IBM J. Res. Dev.},
	volume = {63},
	number = {4/5},
	pages = {9:1-9:12},
	doi = {10.1147/JRD.2019.2915067},
	pagetotal = {1}
}

@dataset{Simbeck.2025,
	title = {Standards for Transparent {{AI}} in {{Human Resource Management}} ({{TRANKI}}) – {{Research}} Data},
	author = {Simbeck, Katharina and Kalff, Yannick},
	date = {2025-11-25},
	publisher = {Zenodo},
	doi = {10.5281/ZENODO.17708792},
	langid = {english}
}

@article{Strohmeier.2009,
	title = {Organizational Adoption of e‐{{HRM}} in {{Europe}}: {{An}} Empirical Exploration of Major Adoption Factors},
	shorttitle = {Organizational Adoption of e‐{{HRM}} in {{Europe}}},
	author = {Strohmeier, Stefan and Kabst, Rüdiger},
	editor = {Gueutal, Hal G.},
	date = {2009-08-14},
	journaltitle = {Journal of Managerial Psychology},
	shortjournal = {J. Manag. Psychol.},
	volume = {24},
	number = {6},
	pages = {482--501},
	doi = {10.1108/02683940910974099},
	langid = {english}
}

@incollection{Torre.2022,
	title = {People {{Analytics}} and {{The Future}} of {{Competitiveness}}: {{Which Capabilities HR Departments Need}} to {{Succeed}} in the “{{Next Normal}}”},
	booktitle = {{{HR Analytics}} and {{Digital HR Practices}}: {{Digitalization}} Post {{COVID-19}}},
	author = {Torre, Teresina and Sarti, Daria and Antonelli, Gilda},
	editor = {Mondal, Subhra R. and Di Virgilio, Francesca and Das, Subhankar},
	date = {2022-01-01T00:00:00},
	series = {Springer {{eBook Collection}}},
	pages = {1--24},
	publisher = {Palgrave Macmillan},
	location = {Singapore},
	doi = {10.1007/978-981-16-7099-2_1},
	isbn = {978-981-16-7099-2}
}

@article{Tsiskaridze.2023,
	title = {Innovating {{HRM Recruitment}}: {{A Comprehensive Review Of AI Deployment}}},
	author = {Tsiskaridze, Rusudan and Reinhold, Karin and Jarvis, Marina},
	date = {2023-01-01T00:00:00},
	journaltitle = {Marketing and Management of Innovations},
	shortjournal = {Mark. Manag. Innov.},
	volume = {14},
	number = {4},
	pages = {239--254},
	doi = {10.21272/mmi.2023.4-18}
}

@article{VanNoordt.2022,
	title = {Exploratory {{Insights}} on {{Artificial Intelligence}} for {{Government}} in {{Europe}}},
	author = {Van Noordt, Colin and Misuraca, Gianluca},
	date = {2022-04},
	journaltitle = {Social Science Computer Review},
	shortjournal = {Soc. Sci. Comput. Rev.},
	volume = {40},
	number = {2},
	pages = {426--444},
	doi = {10.1177/0894439320980449},
	langid = {english}
}

@article{Weiskopf.2023,
	title = {Algorithmic Governmentality and the Space of Ethics: {{Examples}} from ‘{{People Analytics}}’},
	author = {Weiskopf, Richard and Hansen, Hans Kause},
	date = {2023-01-01T00:00:00},
	journaltitle = {Human Relations},
	shortjournal = {Hum. Relat.},
	volume = {76},
	number = {3},
	pages = {483--506},
	publisher = {Sage Publications},
	doi = {10.1177/00187267221075346}
}

@article{Widder.2023,
	title = {Dislocated Accountabilities in the “{{AI}} Supply Chain”: {{Modularity}} and Developers’ Notions of Responsibility},
	shorttitle = {Dislocated Accountabilities in the “{{AI}} Supply Chain”},
	author = {Widder, David Gray and Nafus, Dawn},
	date = {2023-01},
	journaltitle = {Big Data \& Society},
	shortjournal = {Big Data Soc.},
	volume = {10},
	number = {1},
	pages = {20539517231177620},
	doi = {10.1177/20539517231177620},
	langid = {english}
}

@inproceedings{William.2023,
	title = {Enterprise {{Human Resource Management Model By Artificial Intelligence Digital Technology}}},
	booktitle = {2023 4th {{International Conference}} on {{Computation}}, {{Automation}} and {{Knowledge Management}} ({{ICCAKM}})},
	author = {William, P. and Agrawal, Amit and Rawat, Navneet and Shrivastava, Anurag and Srivastava, Arun Partap and {Ashish}},
	date = {2023-12-12},
	pages = {01--06},
	publisher = {IEEE},
	location = {Dubai, United Arab Emirates},
	doi = {10.1109/ICCAKM58659.2023.10449624},
	eventtitle = {2023 4th {{International Conference}} on {{Computation}}, {{Automation}} and {{Knowledge Management}} ({{ICCAKM}})},
	isbn = {979-8-3503-9324-8}
}

@article{Williams.2024,
	title = {Framing {{Algorithmic Management}}: {{Constructed Antagonism}} on {{HR Technology Websites}}},
	author = {Williams, Penny and Khan, Maria Hameed},
	date = {2024-01-01T00:00:00},
	journaltitle = {New Technology, Work and Employment},
	shortjournal = {New Technol. Work Employ.},
	doi = {10.1111/ntwe.12305},
	volumes = {ntwe.12305}
}

@article{Wirges.2023,
	title = {Towards a Process-Oriented {{Understanding}} of {{HR Analytics}}: {{Implementation}} and {{Application}}},
	author = {Wirges, Felix and Neyer, Anne-Katrin},
	date = {2023-01-01T00:00:00},
	journaltitle = {Review of Managerial Science},
	shortjournal = {Rev. Manag. Sci.},
	volume = {17},
	number = {6},
	pages = {2077--2108},
	doi = {10.1007/s11846-022-00574-0}
}

@article{Zhai.2024,
	title = {{{AI}} in {{Human Resource Management}}: {{Literature Review}} and {{Research Implications}}},
	author = {Zhai, Yuming and Zhang, Lixin and Yu, Mingchuan},
	date = {2024-01-01T00:00:00},
	journaltitle = {Journal of the Knowledge Economy},
	shortjournal = {J. Knowl. Econ.},
	doi = {10.1007/s13132-023-01631-z}
}

%\bmsection*{Supporting information}
%Additional supporting information may be found in the online version of the article at the publisher’s website.

%\nocite{*}% Show all bib entries - both cited and uncited; comment this line to view only cited bib entries;

%\bmsection*{Author Biography}
%Redacted for double blind peer review.
%\begin{biography}{\includegraphics[width=76pt,height=76pt,draft]{empty}}{
%{\textbf{Author Name.} Please check with the journal's author guidelines whether
%author biographies are required. They are usually only included for
%review-type articles, and typically require photos and brief
%biographies for each author.}}
%\end{biography}
%
%\begin{biography}{\includegraphics[width=76pt,height=76pt,draft]{empty}}{
%{\textbf{Author Name.} Please check with the journal's author guidelines whether
%author biographies are required. They are usually only included for
%review-type articles, and typically require photos and brief
%biographies for each author.}}
%\end{biography}

\end{document}